\documentclass[fleqn,usenatbib]{mnras}

\usepackage{newtxtext}
\usepackage[T1]{fontenc}
\usepackage{booktabs}

\DeclareRobustCommand{\VAN}[3]{#2}
\let\VANthebibliography\thebibliography
\def\thebibliography{\DeclareRobustCommand{\VAN}[3]{##3}\VANthebibliography}

\usepackage{graphicx}	% Including figure files
\usepackage{amsmath}	% Advanced maths commands
\usepackage{amssymb}	% Extra maths symbols
\usepackage[usenames, dvipsnames]{color}
\usepackage{overpic}

\newcommand{\msun}{\,{\rm M_\odot}}

\newcommand{\beq}{\begin{equation}}
\newcommand{\eeq}{\end{equation}}
\newcommand{\ba}{\begin{eqnarray}}
\newcommand{\ea}{\end{eqnarray}}
\newcommand{\hagn}{\mbox{{\sc \small Horizon-AGN}}}
\newcommand{\fedd}{$f_{\rm Edd}$}

\title[Dual AGN in Horizon-AGN]{Dual AGN in the Horizon-AGN simulation and their link to galaxy and massive black hole mergers, with an excursus on multiple AGN}

\author[Volonteri et al.]{Marta Volonteri,$^{1}$\thanks{E-mail: martav@iap.fr (MV)}
Hugo Pfister,$^{2,3}$
Ricarda Beckmann,$^{4}$
Massimo Dotti,$^{5,6,7}$\newauthor
Yohan Dubois,$^{1}$
Warren Massonneau,$^{1}$
Gibwa Musoke$^{8}$
and Michael Tremmel$^{9}$
\\
$^{1}$Institut d'Astrophysique de Paris, Sorbonne Universit\'e, CNRS, UMR 7095, 98 bis bd Arago, 75014 Paris, France\\
$^{2}$Department of Physics, The University of Hong Kong, Pokfulam Road, Hong Kong, China\\
$^{3}$DARK, Niels Bohr Institute, University of Copenhagen, Jagtvej 128, 2200 København, Denmark\\
$^{4}$Institute of Astronomy and Kavli Institute for Cosmology, University of Cambridge, Madingley Rd, Cambridge CB3 0HA, United Kingdom\\
$^{5}$Dipartimento di Fisica G. Occhialini, Universit\`a di Milano-Bicocca, Piazza della Scienza 3, I-20126 Milano, Italy\\
$^{6}$INAF, Osservatorio Astronomico di Brera, Via E. Bianchi 46, I-23807 Merate, Italy\\
$^{7}$INFN, Sezione di Milano-Bicocca, Piazza della Scienza 3, I-20126 Milano, Italy\\
$^{8}$Anton Pannekoek Institute for Astronomy, University of Amsterdam, Science Park 904, 1098 XH Amsterdam, The Netherlands\\
$^{9}$Astronomy Department, Yale University, P.O. Box 208120, New Haven, CT 06520, USA\\
}

\date{Accepted XXX. Received YYY; in original form ZZZ}

\pubyear{2020}

\begin{document}
\label{firstpage}
\pagerange{\pageref{firstpage}--\pageref{lastpage}}
\maketitle

\begin{abstract}
The occurrence of dual active galactic nuclei (AGN) on scales of a few tens of kpc can be used to study merger-induced accretion on massive black holes (MBHs) and to derive clues on MBH mergers, using dual AGN as a parent population of precursors. We investigate the properties of dual AGN in the cosmological simulation \hagn. We create catalogs of dual AGN selected with distance and luminosity criteria, plus sub-catalogs where further mass cuts are applied. We divide the sample into dual AGN hosted in different galaxies, on the way to a merger, and into those hosted in one galaxy, after the galaxy merger has happened. We find that the relation between MBH and galaxy mass is similar to that of general AGN population and we compare the properties of dual AGN also with a control sample, discussing differences and similarities in masses and Eddington ratios. The typical mass ratio of galaxy mergers associated to dual AGN is 0.2, with mass loss in the smaller galaxy decreasing the mass ratio as the merger progresses. Between 30 and 80 per cent of dual AGN with separations between 4 and 30~kpc can be matched to an ensuing MBH merger. The dual AGN fraction increases with redshift and with separation threshold, although above 50~kpc the increase of multiple AGN limits that of duals.  Multiple AGN are generally associated with massive halos, and mass loss of satellites shapes the galaxy-halo relation.
 
\end{abstract}


%%%%%%%%%%%%%%%%%%%%%%%%%%%%%%%%%%%%%%%%%%%%%%%%%%

%%%%%%%%%%%%%%%%% BODY OF PAPER %%%%%%%%%%%%%%%%%%

\section{Introduction}
Dual active galactic nuclei (AGN), with separations of hundreds of parsecs to tens of kiloparsecs, have received increasing attention, either to study the link between galaxy mergers and massive black hole (MBH) fueling, or as precursors of MBH mergers. The recent review by \cite{2019NewAR..8601525D} summarizes both the theoretical and observational status of the field. 

From the theoretical point of view, after simple early models \citep{2002MNRAS.336L..61H,VHM}, more refined phenomenological \citep{2011ApJ...738...92Y} and numerical investigations \citep{VW2012,2013MNRAS.429.2594B,2016MNRAS.458.1013S,2016MNRAS.460.2979V,2017MNRAS.469.4437C,2019MNRAS.483.2712R,2020MNRAS.492.5620B,2021ApJ...916..110L,2021ApJ...916L..18R} have addressed the occurrence of dual and multiple AGN residing in the same galaxy, or in galaxies separated by up to a few tens of kpc. Studies in idealized set-ups have highlighted that two AGN in merging galaxies do not necessarily light up at the same time \citep{VW2012,2017MNRAS.469.4437C}, and that the mass ratios of merging galaxies \citep{2017MNRAS.469.4437C}, the orbital parameters of MBHs as well as the structure and kinematics of the host galaxy play a role \citep{2021ApJ...916..110L}. Cosmological simulations, which have lower resolution than idealized simulations, have instead focused on the incidence of dual AGN and on their origins. \cite{2016MNRAS.458.1013S} have analyzed the differences between dual and offset AGN. \cite{2016MNRAS.460.2979V} and \cite{2021ApJ...916L..18R} have considered dual AGN in the context of wandering MBHs, the population of MBHs that does not settle in the galaxy center (and therefore is unable to merge with the central MBH). \cite{2019MNRAS.483.2712R} have investigated the abundance of dual AGN as a function of redshift and confirmed that non-simultaneous accretion on MBHs decreases the detection probability. \citet{2020MNRAS.492.5620B} have expanded to multiple AGN, while \cite{2020ApJ...904..150B} have studied the accretion properties of dual and multiple AGN. 

Observationally, many dual AGN have been discovered serendipitously, but systematic searches have started addressing the statistical properties of dual AGN, both their occurrence and properties. Searches are generally of two types, either blind searches that search surveys for two AGN at small separation or in the same galaxy, for instance through spectroscopic signatures \citep{2013ApJ...777...64C,2020ApJ...904...23K,2020ApJ...888...73H}, or assisted searches that look for companions near detected AGN \citep[e.g.,][]{2012ApJ...746L..22K,2020ApJ...899..154S}. After selection of candidates, additional tests are often needed to confirm the dual nature of the selected AGN \citep[e.g.,][]{2011ApJ...739...44R,2019MNRAS.484.4933R,2016ApJ...826..106G,2020ApJ...892...29F,2012ApJ...753...42C}. A small number of multiple AGN have also been reported in the literature \citep[e.g.,][]{2007ApJ...662L...1D,2015MNRAS.453..214D,2015Sci...348..779H,2019ApJ...883..167P,2019ApJ...887...90L}, with separations varying from a few tens to hundreds of kpc.

To ease comparison with systematic searches of dual AGN and other theoretical investigations of dual AGN in a cosmological context, in this paper we use a large cosmological simulation, \hagn, to investigate the properties of dual AGN and their link to galaxy and MBH mergers, addressing the question of whether dual AGN are a good proxy as precursors of MBH mergers. In the search for dual AGN we realized that multiple AGN systems (3 or more AGN) ``pollute'' the dual AGN sample, and therefore separated multiple AGN from ``pure'' dual AGN. This led us to explore the properties and occurrence of multiple AGN, and in particular their environments.

\section{The \hagn~simulation}

The \hagn~simulation is run with the adaptive mesh refinement code {\mbox{{\sc \small RAMSES}}}~\citep{teyssier02}. It covers a large volume, (142 comoving Mpc)$^3$, at a relatively low spatial and mass resolution: cell refinement is permitted down to $\Delta x=1$~kpc, the dark matter particle mass is $8\times 10^7 \msun$, the stellar particle mass is $2\times 10^6 \msun$, and the MBH seed mass is $10^5 \msun$. 

The simulation includes all standard galaxy formation implementations. Gas cooling is modelled using curves from \cite{sutherland&dopita93} down to $10^4\, \rm K$. The gas follows an equation of state for an ideal monoatomic gas with an adiabatic index of $5/3$. A uniform UV background is included after redshift $z_{\rm reion} = 10$ following \cite{haardt&madau96}. Star formation adopts a Schmidt relation with a constant star formation efficiency $\epsilon_*=0.02$ \citep{kennicutt98, krumholz&tan07} in regions which exceed gas hydrogen number density $n_0=0.1\, \rm H\, cm^{-3}$ following a Poisson random process \citep{rasera&teyssier06, dubois&teyssier08winds}. Feedback from Type Ia SNe, Type II SNe and stellar winds is included assuming a \citet{1955ApJ...121..161S} initial mass function with cutoffs at $0.1\, \rm M_{\odot}$  and $100 \, \rm M_{\odot}$. 

MBH formation is based on local gas properties down to $z=1.5$, after which it is stopped. Seeds with mass $10^5\, \rm M_{\odot}$ are created in cells with gas density larger than $n_0$ and gas velocity dispersion  larger than $100\, \rm km\,s^{-1}$. To avoid formation of multiple MBHs in the same galaxy, an exclusion radius of 50 comoving kpc is imposed. The accretion rate adopts a Bondi-Hoyle-Littleton approach, modified by a factor $\alpha=(n/n_0)^2$ when $n>n_0$ and $\alpha=1$ otherwise \citep{Booth2009} in order to account for the inability to capture the multiphase nature of the interstellar gas at these resolutions. The radiative efficiency is fixed at 0.1, and the accretion rate onto MBHs is capped at the Eddington luminosity. AGN feedback takes two forms, thermal at high accretion rated and kinetic otherwise. Above 1 per cent of the Eddington luminosity, 15\% of the MBH emitted luminosity is isotropically coupled to the gas within $4\Delta x$ as thermal energy. Below 1 per cent of the Eddington luminosity, 100 per cent of the power is injected into a bipolar outflow with velocity $10^4\,\rm km\, s^{-1}$, injected in a cylinder with radius $\Delta x$ and height $2 \, \Delta x$. 

MBH dynamics is corrected with an explicit inclusion of drag force from the gas onto the MBH \citep{2012MNRAS.420.2662D}. The magnitude of this force is expressed as $F_{\rm DF}= f_{\rm gas} 4 \pi \alpha \rho_{\rm gas} (G M_{\rm BH}/\bar c_s)^2$, where $\rho_{\rm gas}$ is the mass-weighted mean gas density within a sphere of radius $4 \, \Delta x$, $f_{\rm gas}$ is a factor function of the mach number ${\mathcal M}=\bar u/\bar c_s$ \citep{Ostriker1999}, with  $\bar u$ and $\bar c_s$ the mass-weighted relative speed of the MBH with respect to surrounding gas and sound speed, and $\alpha$ is  the same boost factor used for accretion. See \cite{Dubois2013} for additional details. MBHs are merged when they are separated by $\leqslant 4 \Delta x$, corresponding to 4~kpc, and they are energetically bound in vacuum. 

Dark matter halos and sub-halos are identified with HaloMaker, which uses AdaptaHOP \citep{aubertetal04,Tweed_09}.  A total of 20 neighbours are used to compute the local density of each particle, with a density threshold at 178 times the average total matter density and a threshold of 50 particles. Galaxies are identified in the same way, and they are associated to halos a posteriori, with the main galaxy in a given halo defined as the most massive galaxy within 10 per cent of the halo virial radius.

\begin{figure*}
	\includegraphics[width=0.32\textwidth]{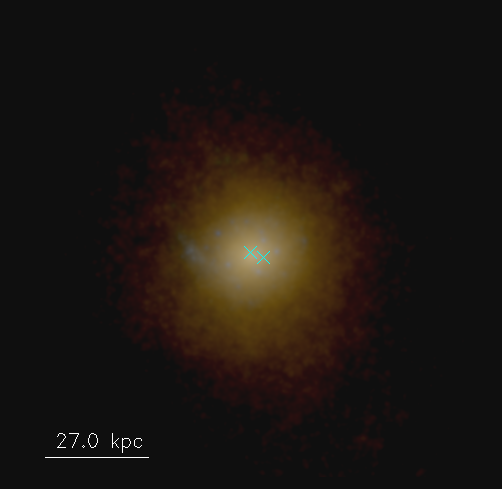}
	\includegraphics[width=0.31\textwidth]{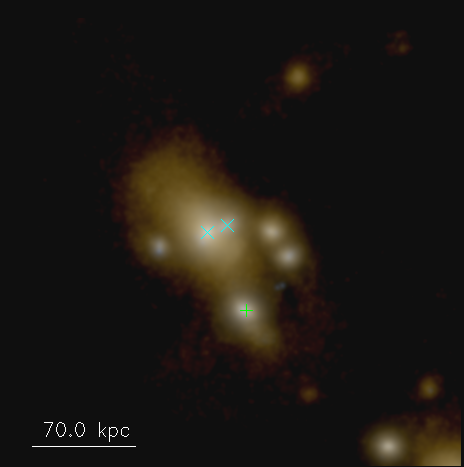}
	\includegraphics[width=0.313\textwidth]{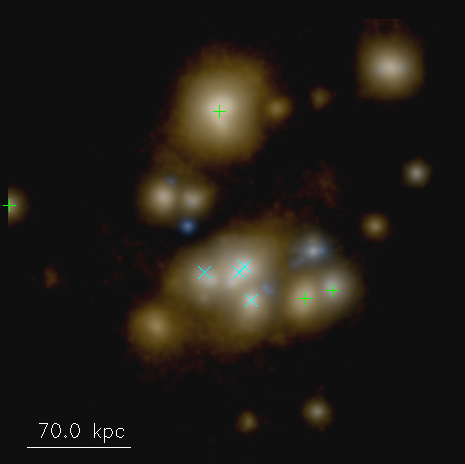}
    \caption{Examples of dual/multiple AGN in false gri colors. Left: a dual AGN in a massive galaxy at $z=0$; middle: a dual AGN at $z=1$; right: a quadruple system at $z=1$. The dual/multiple AGN are shown with crosses, additional AGN in the region with plus signs. In the right panel, the two AGN highlighted with plus signs on the right side are not a dual system: their separation in the direction orthogonal to the plane is 215~kpc.}
    \label{fig:images}
\end{figure*}

\section{Dual AGN catalogues}
\subsection{Selection of dual AGN}
We perform this analysis at 7 outputs, from $z=0$ to $z=3$ in steps of 0.5 in redshift. We refer to these redshifts as $z$, or $z_{obs}$ when we want to stress that this is the redshift/time when the dual is caught in an observation. We first build a catalogue of `central MBHs', defined as the most massive MBH located within 10 per cent of a halo virial radius, and within twice the effective radius, $R_e$ of the most massive galaxy hosted within 10 per cent of that halo \citep[see][for more details]{2016MNRAS.460.2979V}. We then remove the central MBHs from the list of MBHs, and we assign the remaining ones to galaxies, selecting the closest galaxy when a MBH can be associated to multiple galaxies. We aim to include only MBHs that are physically associated with a galaxy, because we want to explore the properties of galaxies hosting dual AGN. We therefore limit our analysis to MBHs within $4\times R_e$ of a galaxy\footnote{For density profiles with slope between $-1$ and $-2$,  $4\times R_e$ corresponds to 2-3 times the radius that contains 90 per cent of the mass, ensuring that all the visible part of galaxies is included.}, which excludes MBHs that are far outside the baryon dominated region. This choice does not strongly affect our selection of AGN, as this only removes 5-7 per cent of MBHs with luminosities greater than $10^{42} \,\rm{erg \, s^{-1}}$ located inside the virial radius of dark matter halos.

We then apply a luminosity cut of $10^{42} \,\rm{erg \, s^{-1}}$ to all MBHs associated with galaxies, and thus define the full sample of AGN studied in this paper. To identify dual and multiple AGN within this sample, we search for AGN located within  30~kpc (physical 3D distance) of each other. The choice of 30~kpc is motivated theoretically by focusing on systems that are or will be involved in an interaction, and observationally by avoiding chance superpositions, while keeping close to the typical separations used in observational searches \citep[e.g.,][]{2013ApJ...777...64C,2020ApJ...899..154S}. Since MBHs are merged when their separation is less than 4~kpc, we cannot track duals closer than that. For some analyses we extend the distance cut to 50~kpc.

 In a pair, the fainter AGN in the pair is referred to as ``secondary'', and its properties are identified with a subscript ``2''. The brighter AGN in the pair, ``primary'' has its properties identified with a subscript ``1''. Primary and secondary are selected at the same output, i.e., at the same redshift. 
For instance, $M_{\rm BH,1}$ is the mass of the MBH powering the more luminous AGN in the pair (the primary), and can be higher or lower than $M_{\rm BH,2}$, the mass of the MBH powering the secondary. The same goes for the Eddington ratios and galaxy stellar masses.  

Given our procedure, two non-central MBHs can be selected as a dual AGN, and multiple AGN systems can be counted as more than one dual. For instance if within a region of space with 30~kpc radius three AGN pass the criteria, three dual AGN are counted and analyzed separately.  To avoid overcounting dual AGN, we therefore proceed hierarchically from multiplets to duals. We first identify clusters of 6 AGN -- the highest multiple for our reference luminosity and distance cuts -- and remove them from the list, we then proceed similarly for quintuplets, quadruplets and triplets and we are left with ``pure'' dual AGN. This procedure is repeated for different luminosity thresholds, since a quadruple system identified using a threshold of $10^{42} \,\rm{erg \, s^{-1}}$ would become a dual system for a  threshold of $10^{44} \,\rm{erg \, s^{-1}}$ if two of the AGN are too faint to be picked up with the high luminosity threshold. We often use the convention of referring to luminosity thresholds as ``$\log(L_{\rm bol})$'', with the bolometric luminosity expressed in $\rm{erg \, s^{-1}}$.

Pairs passing a single luminosity criterion define the general sample: in this case both AGN must pass the same luminosity threshold. We show in  the Appendix how increasing the luminosity threshold or decreasing the distance threshold modify the results. For luminosity thresholds higher than $10^{42} \, \rm{erg \, s}^{-1}$, we also create dual AGN samples relaxing the criterion on the secondary AGN in the pair, under the assumption that the primary has been identified, and a fainter companion is searched for in its surroundings. We show results for secondary AGN having luminosity larger than 1/10 the primary's luminosity, and apply this criterion only to primaries with $\log(L_{\rm bol})>43$.

Additional subsamples can be created by applying further criteria to both all AGN and dual AGN. We analyze a galaxy mass selected sample, for AGN hosted in galaxies with total stellar mass~$>10^{10} \msun$, and a MBH mass selected samples, for AGN powered by MBHs with mass $>10^7 \msun$ (applied to both primary and secondary or only to the primary). Such selections are often used either in  simulations, to ensure that only MBHs in well-resolved galaxies are included, or observations (depending on the parent sample where dual AGN are searched for). We want to explore here possible biases arising from applying such cuts. 

We often divide dual AGN in two groups: those hosted in two different galaxies and those hosted in a single galaxy \citep[see also][]{2019MNRAS.483.2712R}. The reason for differentiating is that they trace physically different stages. The former are either on the way to a galaxy merger or a chance superposition. The latter are the byproduct of a galaxy merger, either an actively decaying MBH towards a MBH pair/binary or one/two wandering MBH whose dynamical evolution is inefficient\footnote{This approach is complementary to \cite{2021ApJ...916L..18R} who select wandering MBHs first and then investigate which ones can be identified as dual AGN.}. Dual AGN hosted in different galaxies outnumber duals hosted in one galaxy by a factor of about $5:1$. 

Visual examples of the dual/multiple AGN in our samples are shown in Fig.~\ref{fig:images}.
In this paper we will use the hierarchically created samples of pure dual AGN and multiplets, generally for a luminosity threshold of $10^{43} \,\rm{erg \, s^{-1}}$ ($\log(L_{\rm bol})>43$), to consider AGN that are sufficiently powerful to be identified observationally, but we provide additional catalogues of pure dual AGN and basic catalogues of dual AGN regardless of multiplicity for a variety of distance (between 10 and 50~kpc) and luminosity thresholds (from $\log(L_{\rm bol})>40$ to $\log(L_{\rm bol})>44$). See Data Availability.

\subsection{Linking dual AGN to galaxy mergers}
\label{sec:LinkingDualAGNToGalaxyMergers}
Dual AGN have been proposed as signposts of galaxy mergers \citep{2009ApJ...698..956C}. To test this hypothesis we trace if a dual AGN observed at a given time can be matched with a preceding or ensuing galaxy merger. 

For dual AGN hosted in different galaxies at some observation redshift $z$, we use the galaxy merger tree (obtained with \textsc{TreeMaker}, \citealt{Tweed_09}) to obtain the list of main descendants for the primary and secondary galaxies. We define the main descendant of a galaxy in output $i$ the galaxy in output $i+1$ which shares most mass with the progenitor galaxy. If the two galaxies merge, their descendants become identical at some point, and we denote by $z_{\rm galmerg}$ the redshift at which this  happens. Since we are interested in the relation between galaxy mergers and dual AGN, we do not consider for this analysis pairs in galaxies which have not merged by $z=0$, which represent only 1.8 per cent of duals at $\log(L_{\rm bol})>43$.

For dual AGN that are hosted in the same galaxy at redshift $z$, we search back in time to find an output where the MBHs were hosted in different galaxies and we then apply the exact same strategy as above to obtain the redshift of the galaxy merger $z_{\rm galmerg}$. This part of the analysis encountered two difficulties. First, sometimes no separate host galaxy can be found for the two MBHs, despite the criterion of a minimum distance to any existing MBH for formation of another MBH. The reason is related to MBHs forming with a criterion based only on gas properties and on MBHs being free to move. Sometimes MBHs form in a gas cloud that is not associated to a galaxy, because it does not pass the criterion for the halo/galaxy finder. Later on one of these intergalactic MBHs may get captured by a halo/galaxy and at that point, if some stochastic accretion occurs, it can be picked up as a member of a dual AGN system. This occurs mainly in two cases: (i) either the intergalactic MBH is formed at very high redshift (identified at the first output or rarely at the second output) and wanders outside any identified galaxy for a very long time or (ii) the MBH forms shortly before the output where dual AGN are selected. The number of these cases decreases with redshift because gas density decreases as well, therefore it is harder to form a MBH, and for a wandering MBH it is more difficult to accrete from the host. At $\log(L_{\rm bol})>43$ these cases represent about 30 percent of dual AGN hosted in the same galaxy at $z=3$ (and 3 percent of all dual AGN), dropping after MBH formation has been stopped.  The second difficulty is that we expect to have $z_{\rm galmerg}>z$: the MBHs are identified in the same galaxy, therefore the galaxy merger should have happened earlier. However, around the time of a galaxy merger a MBH can sometimes be associated to two galaxies, since the galaxies spatially overlap and a MBH could be located in that region. When a MBH can be associated to multiple galaxies, we pick the galaxy the MBH is closer to, but we could have picked the other one. In practice such dual AGN could have been equally been categorized as "in the same galaxy" or "in different galaxies". This explains why for some dual AGN hosted "in the same galaxy" $z_{\rm galmerg}<z$. These represent only 0.65 per cent of the dual AGN with $\log(L_{\rm bol})>43$.

\subsection{Linking dual AGN to MBH mergers}
\cite{2020MNRAS.498.2219V} have investigated MBH mergers in the \hagn~simulation. We use their results to match dual AGN with MBH mergers and probe whether dual AGN are good predictors for MBH mergers. We consider ``numerical mergers'' where the MBHs are merged in the simulation (meeting the criteria that the separation is $\leq 4\Delta x$ and they are energetically bound in vacuum) and the merger occurs within twice the effective radius of the host galaxy. We also consider the subset of these events where, including post-processed delays to account for the orbital decay from $4\Delta x$ to coalescence, the MBH mergers occurs by $z=0$ (``delayed mergers'').  Since in \hagn~the IDs of MBHs are conserved, we simply look for MBH mergers for which the two MBHs have the same IDs as the MBHs in a dual AGN.  

\section{Properties of dual AGN}

\begin{figure}
	\includegraphics[width=\columnwidth]{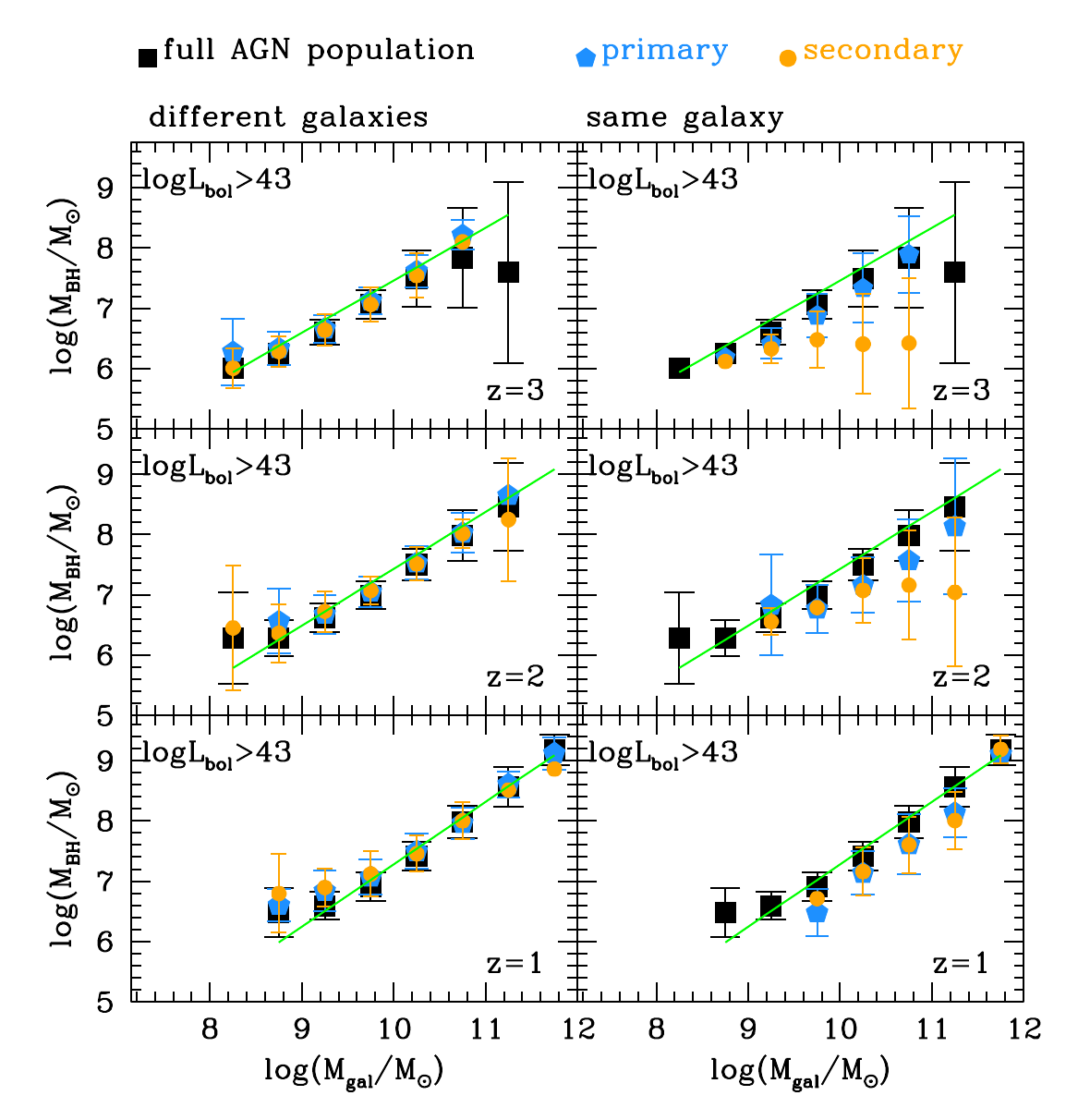}
    \caption{Mean MBH mass in bins of total galaxy stellar mass for all AGN and primaries/secondaries in dual AGN systems, dividing the sample in duals hosted in one or two galaxies. Errorbars show the variance. Mean and variance are calculated in log-space. Dual AGN follow the same relation as the general population, except for some low-mass MBHs in massive galaxies. These MBHs are generally not the central MBHs in those galaxies and appear therefore as the secondary AGN in a dual system hosted in a single galaxy. The solid green line shows the relation for all MBHs (active and inactive) at the same redshift.
    }
    \label{fig:mbh_mgal}
\end{figure}

\begin{figure}
	\includegraphics[width=\columnwidth]{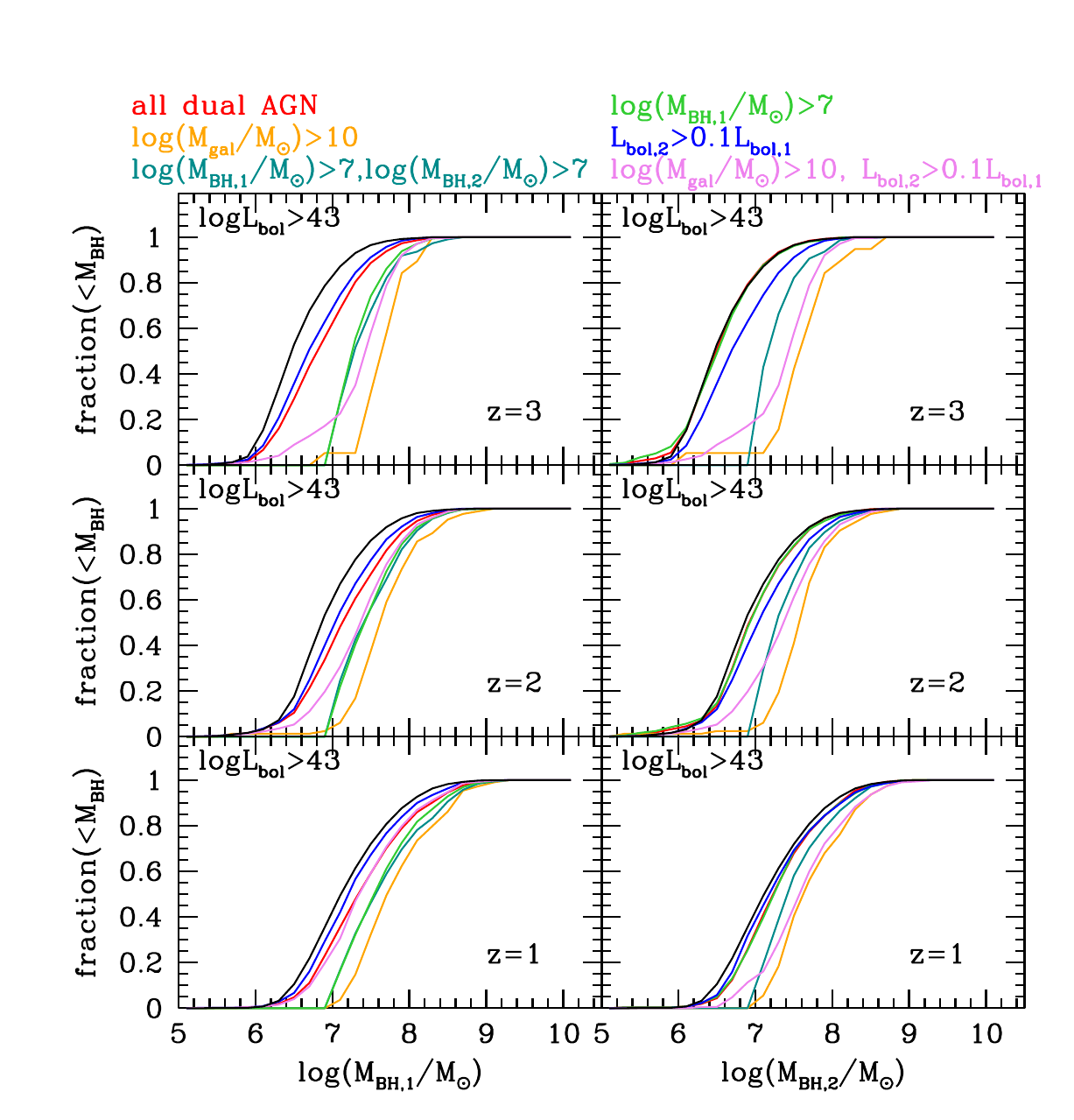}
    	\includegraphics[width=\columnwidth]{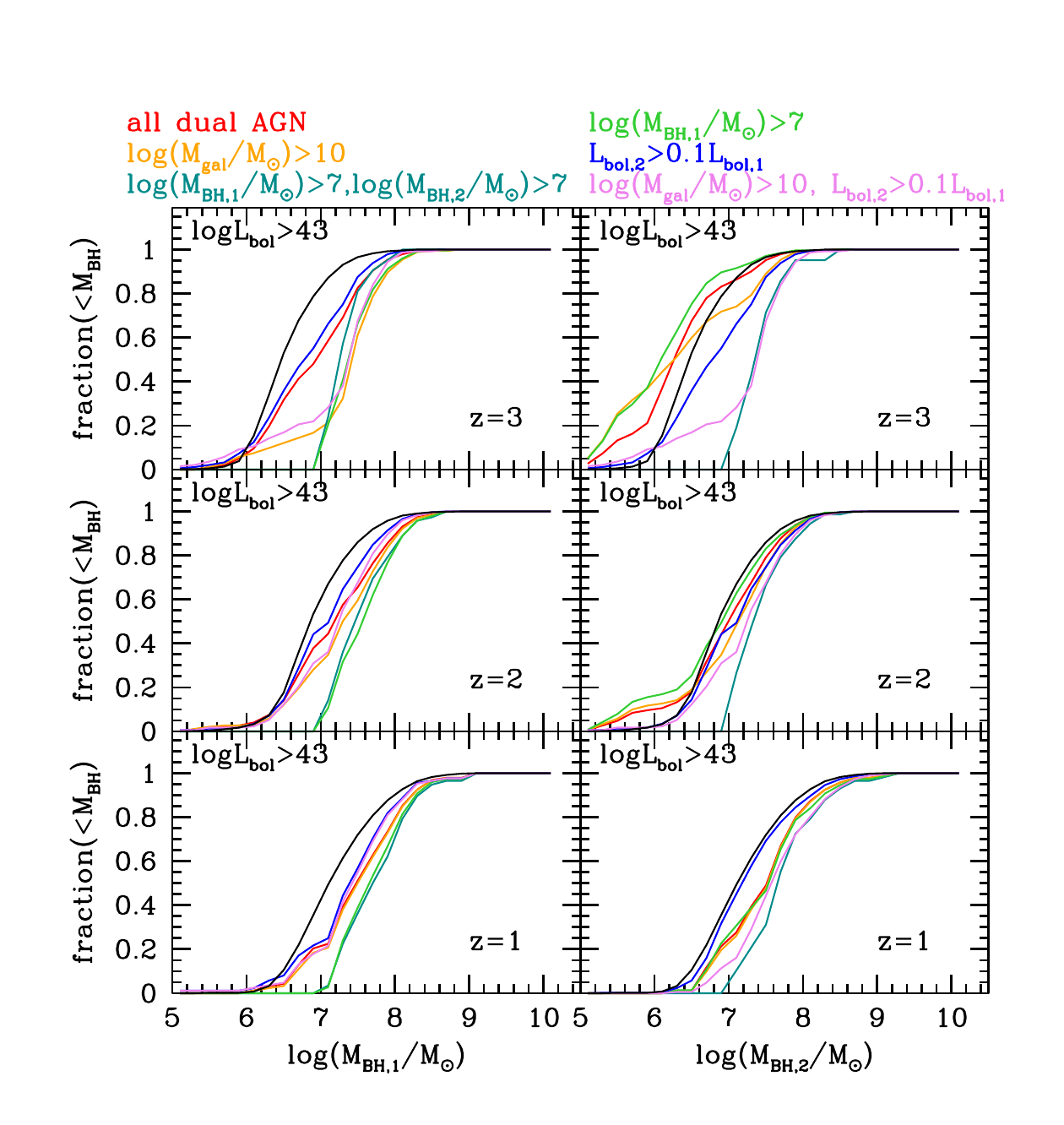}
\caption{Cumulative BH mass distribution for all AGN (black, imposing only a luminosity threshold) and for dual AGN passing some cuts. Left: most luminous AGN in the pair, primary). Right: least luminous AGN in the pair (secondary). Duals hosted in different galaxies are shown in the top 6 panels, duals hosted in one galaxy in the bottom six.}
    \label{fig:mbh_distr}
\end{figure}

\begin{figure}
	\includegraphics[width=\columnwidth]{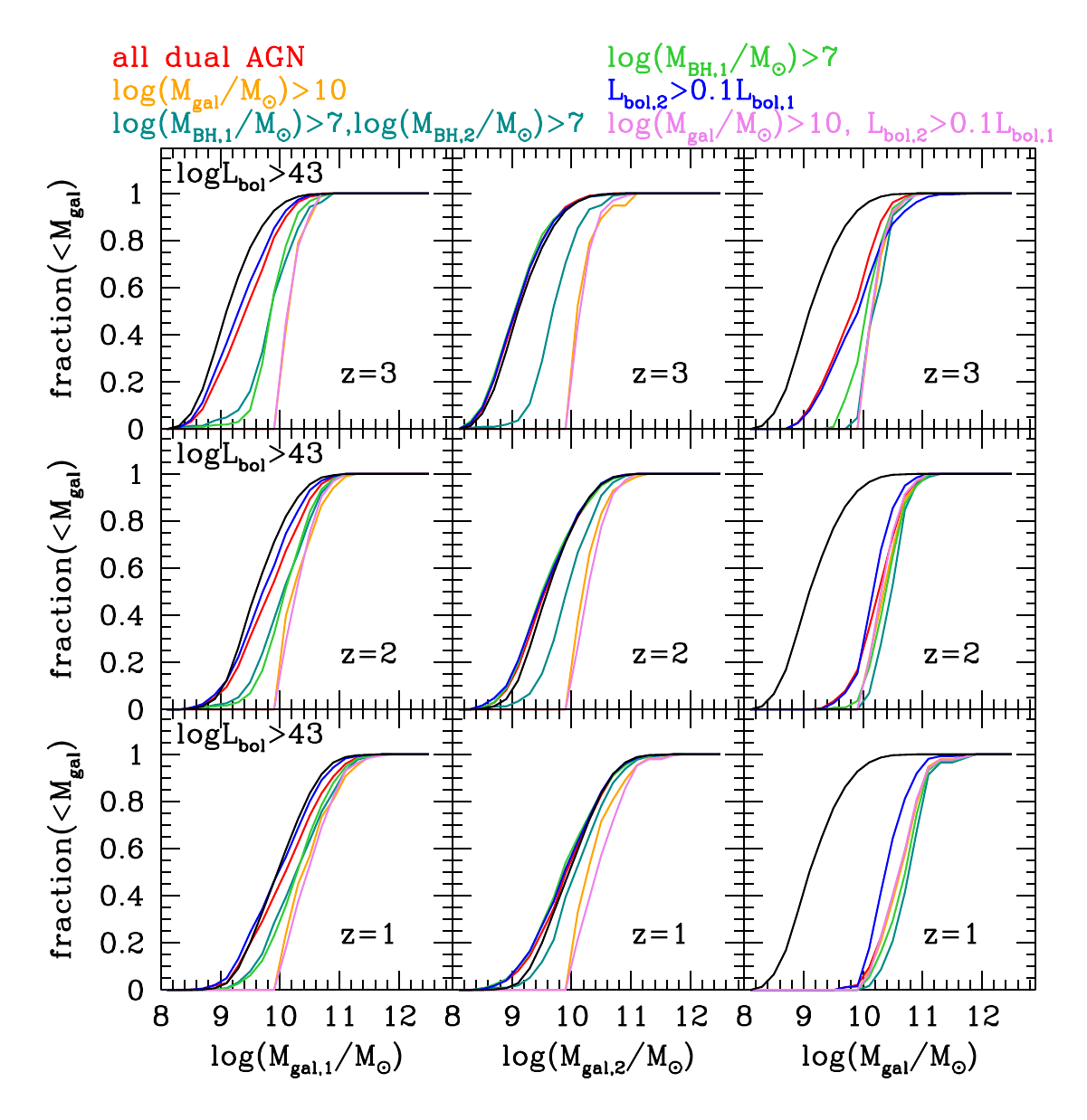}
\caption{Cumulative galaxy mass distribution for all AGN (black, imposing only a luminosity threshold) and for dual AGN passing some cuts.  Duals hosted in different galaxies are shown in the left and middle panels, duals in the same galaxy in the right panel. Here for duals hosted in one galaxy, $M_{\rm gal,1}=M_{\rm gal,2}$ (note that duals in the same galaxy are about 15 per cent of all dual AGN: the sample is dominated by duals hosted in different galaxies). Left: most luminous AGN in the pair, primary). Right: least luminous AGN in the pair (secondary).
}
    \label{fig:mgal_distr}
\end{figure}

\subsection{MBH and galaxy masses}

The relation between MBH and total galaxy mass for all AGN and dual AGN is shown in Fig.~\ref{fig:mbh_mgal} \citep[see][for a discussion on this relation on the whole MBH population in \hagn, and a comparison with observations]{2016MNRAS.460.2979V}. Deviations from the relation are observed only for dual AGN hosted in the same galaxy: generally the secondary AGN is powered, in massive galaxies, by a MBH much less massive than the central MBH. The primary AGN for duals in one galaxy is also somewhat less massive than expected from the relation defined by the full population (although within the scatter). These primary MBHs appear to be growing to ``catch up'' with their host galaxies following a galaxy merger, in agreement with the results from isolated merger simulations \citep{2015MNRAS.447.2123C} and smaller samples in cosmological simulations \citep{2016MNRAS.458.1013S}. This population, i.e.,  dual AGN hosted in the same galaxy, avoids the least massive galaxies at each redshift, as evident from the right panel of Fig.~\ref{fig:mbh_mgal}, suggesting that,  to host two sufficiently luminous AGN, galaxies must have experienced at least one relatively major merger, and that massive galaxies experience, overall, more mergers than their lighter counterpart~\citep[e.g.][]{RodriguezGomez2015,Dubois2016}. The lack of dual AGN in the most massive galaxies at $z=3$ is due to the small number of galaxies in the highest mass bin, and the overall low incidence of dual AGN within the total population of AGN across all mass bins. 

We show cumulative distributions in Figures~\ref{fig:mbh_distr} and~\ref{fig:mgal_distr}.
Qualitatively, the masses of the MBHs powering the primary AGN in a dual systems appear larger than those of the full AGN population selected above the same luminosity. This difference is of course amplified when selecting dual AGN above certain galaxy or MBH masses.

The secondary AGN MBHs appear qualitatively consistent with the general AGN population for duals in different galaxies unless mass cuts are applied to the secondary MBH mass, or if the selection allows for a fainter luminosity for the secondary AGN. In this cases the distribution becomes more complex. For duals hosted in one galaxy, non-central low mass MBHs with relatively high accretion rates become an important sub-population at $z>2$.  This disappears at low redshift because the accretion rate on MBHs decreases overall, and secondary MBHs become instead more massive than the general population. We see the same trends in the galaxy mass distributions: this is because most MBHs sit on a correlation between MBH and galaxy mass, except for non-central high-accretion low mass MBHs in massive galaxies. The right panel of Fig.~\ref{fig:mgal_distr} highlights the preference for duals hosted in one galaxy to inhabit the most massive galaxies at that redshift.

The main conclusion from this qualitative analysis is that, at fixed luminosity, primary AGN are powered by MBHs that are more massive than the full population, in agreement with \cite{2019MNRAS.483.2712R}, and that their host galaxies are also more massive than those hosting AGN of the same luminosity. However, dual AGN sit on the same relation between MBH and galaxy mass: dual AGN are simply generally hosted in massive galaxies, especially when we consider duals hosted in the same galaxy. Results from observations seem  to show that dual quasars more luminous than those analysed here are powered by MBHs that are more massive at fixed stellar mass than the $z=0$ relation, but they inhabit the same region as single quasars \citep{2021ApJ...922...83T}. 

\begin{figure}
	\includegraphics[width=\columnwidth]{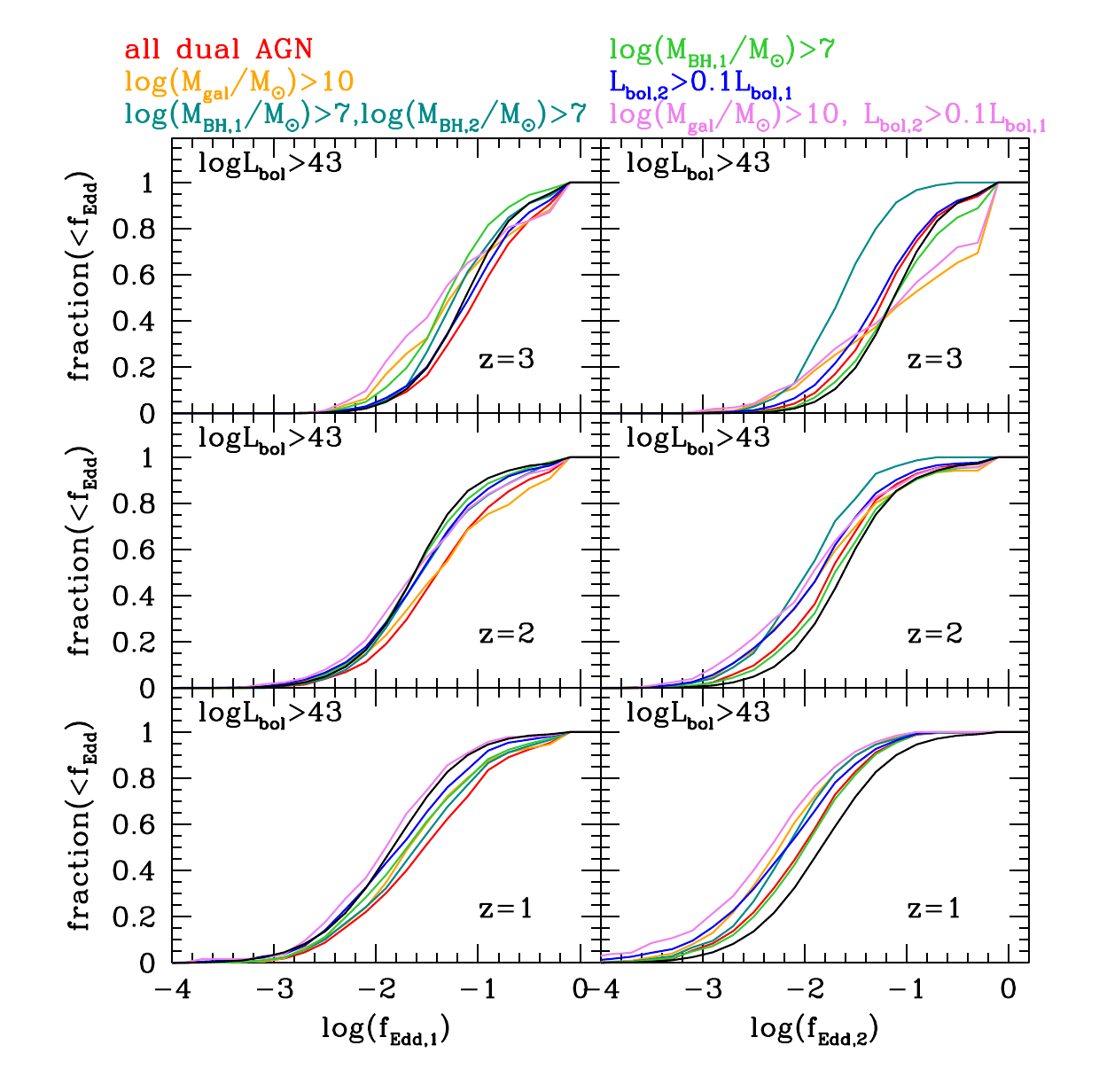}
    \caption{Cumulative Eddington ratio distribution for all AGN (black, imposing only a luminosity threshold) and for dual AGN passing some cuts. Left: most luminous AGN in the pair (primary). Right: least luminous AGN in the pair (secondary).  The primary MBHs of dual AGN have slightly higher accretion rates than the general population, while the MBH in the secondary AGN have slightly lower accretion rates. Imposing mass (black hole or galaxy) cuts can alter significantly the distribution of secondary AGN, and less that of the primary. See text for details.}
    \label{fig:fedd_distr}
\end{figure}

\begin{figure}
	\includegraphics[width=\columnwidth]{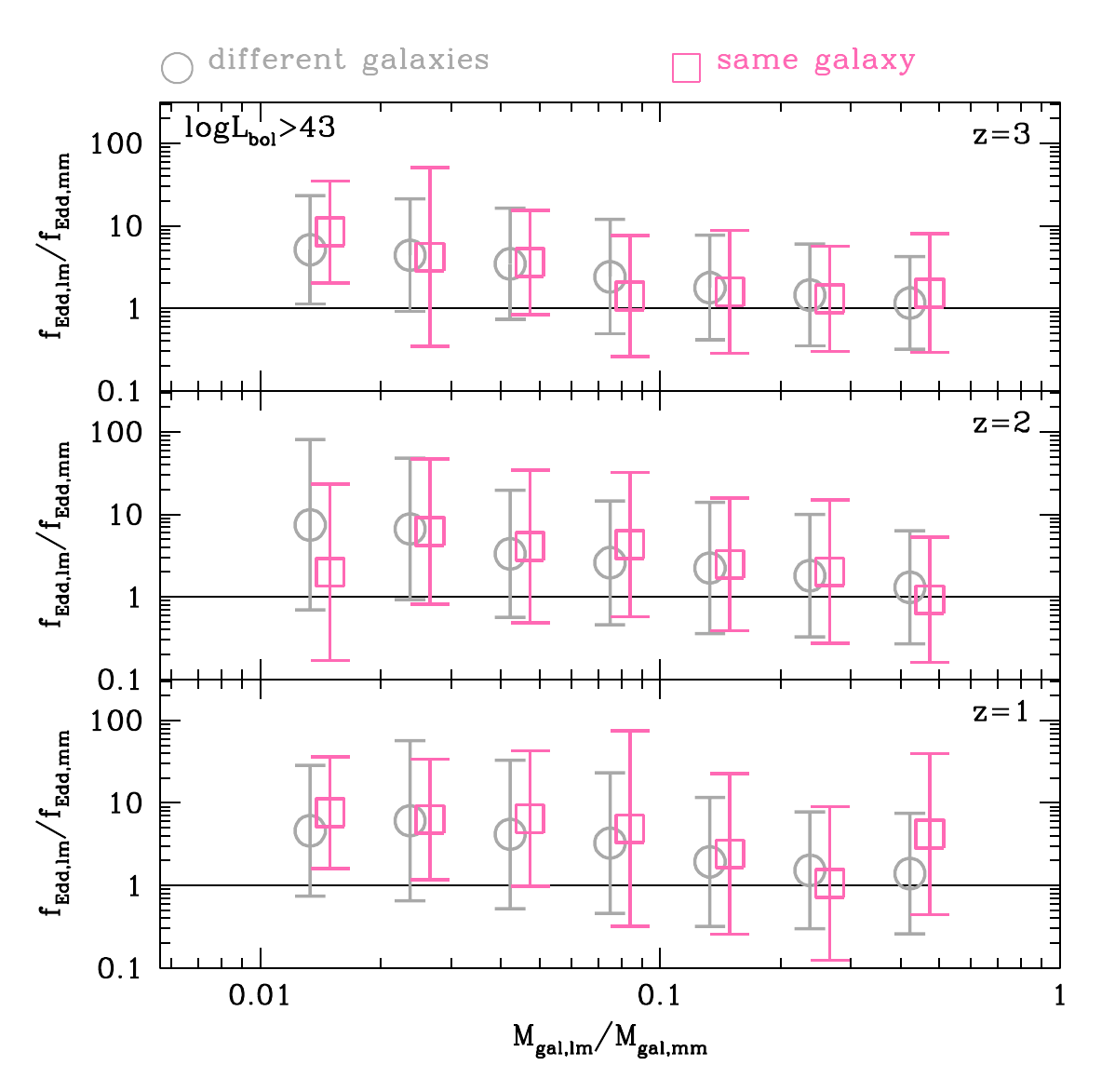}
    \caption{Circles: ratios of Eddington fractions as a function of the ratios of galaxy masses for dual AGN hosted in different galaxies, shifted by 0.05 in the x-axis to improve readability. Squares: for dual AGN hosted in the same galaxy we show the galaxy mass ratio at the time of their merger. The mean is shown with its variance. The subscript `mm' refers to the most massive of the two galaxies, and `lm' to the least massive. Redshifts and luminosity thresholds as reported in the figure.  The MBH in the smaller galaxy has higher \fedd~than the MBH in the larger galaxy, with no significant dependence on redshift.}
    \label{fig:fedd_q}
\end{figure}

\begin{figure}
	\includegraphics[width=\columnwidth]{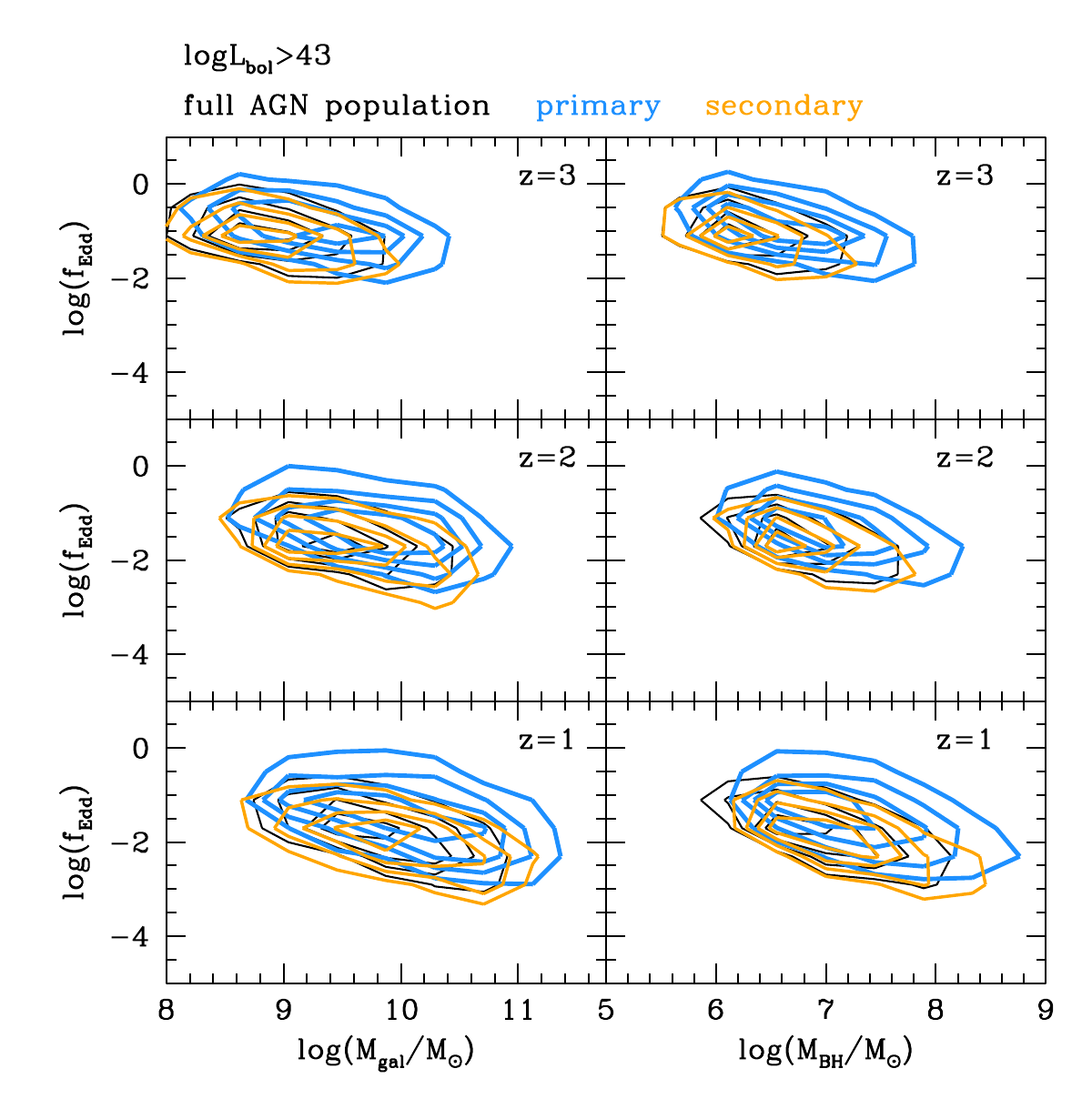}
\caption{Eddington ratio as a function of galaxy and MBH mass for all AGN (thin black  contours), and for primary (thick blue contours) and secondary (medium thickness orange contours) AGN. For each population, we show four linearly spaced contours, containing 20, 40, 60 and 80 per cent of the sample respectively. The most massive MBHs are accreting at the lowest rates. Primary AGN in duals are accreting at higher rates, and preferentially reside in more massive galaxies, than the general AGN population, covering a range that is not common among all AGN.}
    \label{fig:fedd_mass}
\end{figure}

\subsection{Accretion rates}
We examine the distribution of Eddington ratios (\fedd) of primary and secondary AGN in Fig.~\ref{fig:fedd_distr}. If we apply the same luminosity cut to all AGN, the distribution of \fedd~for primary AGN is similar to that of the full AGN sample, with only a slight tendency to higher \fedd~for the primaries as redshift decreases.  Applying mass cuts to the MBH or galaxy does not change the situation dramatically, although at high redshift the primaries in dual AGN  appear somewhat weaker than the full AGN sample. The reason is simply that applying a mass cut in the presence of a luminosity threshold removes from the sample low-mass highly accreting MBHs. This will appear more strongly when we discuss secondary AGN. 

\begin{table}
\begin{tabular}{|l|l|l|l|l|l|l|}
\hline
       & mean  & stdev & mean  & stdev & mean   & stdev \\ \hline
       & $z=3$   & $z=3$      & $z=2$   & $z=2$      & $z=1$    & $z=1$      \\ \hline
$f_{\rm Edd,1d}$  & 0.224 & 0.264    & 0.143 & 0.240    & 0.116  & 0.210    \\ 
$f_{\rm Edd,1c} $& 0.218 & 0.243    & 0.109 & 0.205   & 0.067  & 0.132    \\ 
$f_{\rm Edd,2d} $ & 0.156 & 0.229    & 0.069 & 0.170    & 0.026  & 0.064    \\ 
$f_{\rm Edd,2c}$& 0.114 & 0.148   & 0.043 & 0.096    & 0.027  & 0.056   \\ 
$f_{\rm Edd,a} $ & 0.164 & 0.206    & 0.078 & 0.166    & 0.048  & 0.107    \\ \hline
$\log(M_{\rm BH,1d})$  & 6.949 & 0.547    & 7.261 & 0.596    & 7.510  & 0.612    \\ 
$\log(M_{\rm BH,1c})$ & 6.707 & 0.484    & 7.089 & 0.550    & 7.735  & 0.603    \\ 
$\log(M_{\rm BH,2d})$ & 6.604 & 0.516    & 7.065 & 0.574    & 7.413  & 0.580    \\ 
$\log(M_{\rm BH,2c})$ & 6.520 & 0.406    & 6.936 & 0.458    & 7.167  & 0.542   \\ 
$\log(M_{\rm BH,a})$   & 6.643 & 0.465    & 7.030 & 0.512    & 7.282  & 0.576    \\ \hline
$\log(M_{\rm gal,1d})$  & 9.566 & 0.532    & 9.990 & 0.555    & 10.252 & 0.628    \\ 
$\log(M_{\rm gal,1c})$  & 9.367 & 0.488    & 9.834 & 0.504    & 10.178 & 0.540    \\ 
$\log(M_{\rm gal,2d})$  & 9.300 & 0.527    & 9.804 & 0.567   & 10.122 & 0.626    \\ 
$\log(M_{\rm gal,2c})$ & 9.134 & 0.429    & 9.630 & 0.441    & 9.970  & 0.499    \\ 
$\log(M_{\rm gal,a})$    & 9.255 & 0.471    & 9.637 & 0.449    & 10.072 & 0.510    \\ \hline
\end{tabular}
\caption{Mean and 1$\sigma$ variance for Eddington ratios, galaxy and MBH masses (masses are expressed in solar masses inside the logarithm) for the dual sample (subscript `d'), the control sample (subscript `c') and the full AGN population (subscript `a') at $z=1,2,3$.}
\label{tab:stats}
\end{table}

Secondary AGN have a more varied behaviour. First, they are generally slightly weaker accretors compared to the whole AGN sample (of which dual AGN are a subsample). Note that in this paper we define primary and secondary based on the AGN luminosity, not on the mass of the galaxies hosting the AGN, therefore this is not in disagreement with \cite{2015MNRAS.447.2123C} and \cite{2016MNRAS.458.1013S}. We indeed confirm here on a larger sample the anti-correlated behaviour of \fedd~and mass of the host galaxies found in previous studies, as shown\footnote{For dual AGN hosted in the same galaxy at the time of observation, we show the galaxy mass ratio at the time of their merger, therefore \fedd~and the galaxy mass are not measured at the same time.} in Fig.~\ref{fig:fedd_q}. The MBH in the least massive galaxy has a higher Eddington ratio, but being generally less massive the combination is such that the ratio of luminosities does not scale clearly with the ratio of galaxy masses.   The differences with the general population are not large, and for instance \cite{2021ApJ...922...83T} find  dual quasars inhabit the same (broad) region as single quasars. We expand on quantitative differences in the next section.

Imposing mass cuts alters the distribution in Fig.~\ref{fig:fedd_distr} substantially. Requiring both MBHs to be heavier than $10^7 \msun$ pushes the distribution to lower \fedd, because the Eddington ratio decreases as MBH mass increases (see Fig.~\ref{fig:fedd_mass}) and many secondaries have low mass and high \fedd. The change in shape when imposing a galaxy mass cut is caused by a combination of the same effect with the presence of low-mass MBHs with high accretion rates in high-mass galaxies: generally these are not the central MBHs and they appear mostly at high-redshift since galaxies are gas-rich and significant accretion can occur also in the non-central region. Relaxing the condition on the luminosity of the secondary cures somewhat these changes in the distribution at high redshift. At low redshift the trends remain but the behaviour is less extreme.  

In summary, the primary AGN accretion properties are generally consistent with the full AGN population, although accreting at slightly higher rates, and robust to selection criteria. Secondary AGN are accreting at similar or slightly lower rates compared to the full AGN population, and imposing mass cuts on the MBH or the galaxy exacerbates this difference. However, in order to ensure consistency, applying such cuts when comparing different samples and/or theoretical models would be beneficial to limit biases.

\begin{table*}
\begin{tabular}{|lllll|l|l|l|l|l|l|}
\hline
\multicolumn{1}{|l|}{}                                    & \multicolumn{1}{l|}{$\mathcal{P}$,c} & \multicolumn{1}{l|}{$\mathcal{P}$,a} & \multicolumn{1}{l|}{$N_{d,c}(1\sigma)$} &   StDev     & $N_{d,c}(3\sigma)$ &  StDev      & $N_{d,a}(1\sigma)$ &  StDev     & $N_{d,a}(3\sigma)$ &    StDev   \\ \hline
\multicolumn{5}{|l|}{1D,$z=3$:  $N_d$=$N_c$=2739, $N_a$=50020}                                                                                                                             &                    &        &                    &       &                    &       \\ \hline
\multicolumn{1}{|l|}{$f_{\rm Edd,1}$}                     & \multicolumn{1}{l|}{6.067E-04}       & \multicolumn{1}{l|}{5.192E-29}       & \multicolumn{1}{l|}{275.40}             & 86.98  & 2104.80            & 378.79 & 62.30              & 1.57  & 601.90             & 6.85  \\ 
\multicolumn{1}{|l|}{$f_{\rm Edd,2}$}                     & \multicolumn{1}{l|}{1.960E-06}       & \multicolumn{1}{l|}{2.888E-20}       & \multicolumn{1}{l|}{139.90}             & 19.91  & 1244.60            & 174.58 & 81.00              & 2.26  & 847.10             & 5.07  \\ 
\multicolumn{1}{|l|}{$\log(M_{\rm BH,1})$}                & \multicolumn{1}{l|}{1.988E-43}       & \multicolumn{1}{l|}{7.185E-143}      & \multicolumn{1}{l|}{18.80}              & 1.69   & 176.70             & 13.51  & 12.30              & 0.48  & 112.20             & 2.25  \\ 
\multicolumn{1}{|l|}{$\log(M_{\rm BH,2})$}                & \multicolumn{1}{l|}{1.086E-14}       & \multicolumn{1}{l|}{2.342E-04}       & \multicolumn{1}{l|}{62.00}              & 7.39   & 602.40             & 68.76  & 568.60             & 11.16 & \textgreater{}2739 & 0.00  \\ 
\multicolumn{1}{|l|}{$\log(M_{\rm gal,1})$}               & \multicolumn{1}{l|}{7.724E-35}       & \multicolumn{1}{l|}{2.797E-157}      & \multicolumn{1}{l|}{25.20}              & 2.30   & 254.40             & 8.28   & 11.60              & 0.52  & 105.00             & 2.00  \\ 
\multicolumn{1}{|l|}{$\log(M_{\rm gal,2})$}               & \multicolumn{1}{l|}{1.753E-24}       & \multicolumn{1}{l|}{7.070E-05}       & \multicolumn{1}{l|}{35.40}              & 2.27   & 344.70             & 17.48  & 301.80             & 12.34 & \textgreater{}2739 & 0.00  \\ \hline
\multicolumn{5}{|l|}{1D,$z=2$:  $N_d$=$N_c$=1310, $N_a$=32001}                                                                                                                             &                    &        &                    &       &                    &       \\ \hline
\multicolumn{1}{|l|}{$f_{\rm Edd,1}$}                     & \multicolumn{1}{l|}{3.190E-05}       & \multicolumn{1}{l|}{8.147E-44}       & \multicolumn{1}{l|}{101.40}             & 21.30  & 787.50             & 126.52 & 21.50              & 1.08  & 201.60             & 3.24  \\ 
\multicolumn{1}{|l|}{$f_{\rm Edd,2}$}                     & \multicolumn{1}{l|}{2.060E-02}       & \multicolumn{1}{l|}{8.175E-14}       & \multicolumn{1}{l|}{188.90}             & 54.70  & N/A                & N/A    & 65.70              & 2.00  & 651.00             & 3.16  \\ 
\multicolumn{1}{|l|}{$\log(M_{\rm BH,1})$}                & \multicolumn{1}{l|}{2.953E-12}       & \multicolumn{1}{l|}{1.631E-44}       & \multicolumn{1}{l|}{34.20}              & 5.71   & 315.80             & 51.77  & 20.30              & 1.06  & 188.40             & 4.55  \\ 
\multicolumn{1}{|l|}{$\log(M_{\rm BH,2})$}                & \multicolumn{1}{l|}{2.140E-11}       & \multicolumn{1}{l|}{5.112E-05}       & \multicolumn{1}{l|}{35.40}              & 6.95   & 330.60             & 55.82  & 182.50             & 4.28  & \textgreater{}1310 & 0.00  \\ 
\multicolumn{1}{|l|}{$\log(M_{\rm gal,1})$}               & \multicolumn{1}{l|}{9.566E-13}       & \multicolumn{1}{l|}{7.557E-63}       & \multicolumn{1}{l|}{35.20}              & 4.59   & 326.30             & 38.87  & 15.60              & 0.70  & 140.20             & 3.12  \\ 
\multicolumn{1}{|l|}{$\log(M_{\rm gal,2})$}               & \multicolumn{1}{l|}{2.396E-15}       & \multicolumn{1}{l|}{5.884E-12}       & \multicolumn{1}{l|}{28.50}              & 4.09   & 248.20             & 33.31  & 81.50              & 2.01  & 764.40             & 5.27  \\ \hline
\multicolumn{5}{|l|}{1D,$z=1$:  $N_d$=$N_c$=626, $N_a$=23252}                                                                                                                              &                    &        &                    &       &                    &       \\ \hline
\multicolumn{1}{|l|}{$f_{\rm Edd,1}$}                     & \multicolumn{1}{l|}{9.456E-05}       & \multicolumn{1}{l|}{1.190E-25}       & \multicolumn{1}{l|}{55.10}              & 11.52  & 405.70             & 70.32  & 17.40              & 0.97  & 157.90             & 2.23  \\ 
\multicolumn{1}{|l|}{$f_{\rm Edd,2}$}                     & \multicolumn{1}{l|}{1.905E-01}       & \multicolumn{1}{l|}{4.654E-11}       & \multicolumn{1}{l|}{289.20}             & 212.26 & N/A                & N/A    & 38.10              & 1.73  & 368.60             & 3.95  \\ 
\multicolumn{1}{|l|}{$\log(M_{\rm BH,1})$}                & \multicolumn{1}{l|}{2.897E-03}       & \multicolumn{1}{l|}{1.473E-15}       & \multicolumn{1}{l|}{80.10}              & 20.63  & 496.90             & 181.77 & 29.00              & 1.25  & 278.70             & 3.50  \\ 
\multicolumn{1}{|l|}{$\log(M_{\rm BH,2})$}                & \multicolumn{1}{l|}{1.973E-09}       & \multicolumn{1}{l|}{1.924E-07}       & \multicolumn{1}{l|}{21.00}              & 2.58   & 207.40             & 16.79  & 61.00              & 1.76  & 547.70             & 6.04  \\ 
\multicolumn{1}{|l|}{$\log(M_{\rm gal,1})$}               & \multicolumn{1}{l|}{2.014E-02}       & \multicolumn{1}{l|}{7.084E-16}       & \multicolumn{1}{l|}{115.80}             & 49.80  & N/A                & N/A    & 29.80              & 1.40  & 267.20             & 2.62  \\ 
\multicolumn{1}{|l|}{$\log(M_{\rm gal,2})$}               & \multicolumn{1}{l|}{2.742E-06}       & \multicolumn{1}{l|}{2.945E-04}       & \multicolumn{1}{l|}{34.30}              & 5.77   & 291.80             & 46.20  & 126.10             & 3.41  & \textgreater{}626  & 0.00  \\ \hline
\multicolumn{1}{|l|}{}                                    & \multicolumn{1}{l|}{$\mathcal{P}$,c} & \multicolumn{1}{l|}{$\mathcal{P}$,a} & \multicolumn{1}{l|}{$N_{d,c}(1\sigma)$} &   StDev     & $N_{d,c}(0.99)$ &  StDev      & $N_{d,a}(1\sigma)$ &  StDev     & $N_{d,a}(0.99)$ &    StDev   \\ \hline
\multicolumn{5}{|l|}{2D,$z=3$:  $N_d$=$N_c$=2739, $N_a$=50020}                                                                                                                             &                    &        &                    &       &                    &       \\ \hline
\multicolumn{1}{|l|}{$f_{\rm Edd,1},\log(M_{\rm BH,1})$}  & \multicolumn{1}{l|}{7.788E-35}       & \multicolumn{1}{l|}{4.276E-145}      & \multicolumn{1}{l|}{10.40}              & 1.07   & 136.30             & 8.93   & 5.90               & 0.57  & 62.80              & 1.62  \\ 
\multicolumn{1}{|l|}{$f_{\rm Edd,2},\log(M_{\rm BH,2})$}  & \multicolumn{1}{l|}{1.315E-22}       & \multicolumn{1}{l|}{9.519E-28}       & \multicolumn{1}{l|}{28.00}              & 3.06   & 255.30             & 21.54  & 33.00              & 1.70  & 408.60             & 3.63 \\ 
\multicolumn{1}{|l|}{$f_{\rm Edd,1},\log(M_{\rm gal,1})$} & \multicolumn{1}{l|}{2.551E-22}       & \multicolumn{1}{l|}{1.658E-140}      & \multicolumn{1}{l|}{16.70}              & 2.11   & 216.80             & 23.23  & 5.60               & 0.52  & 65.00              & 1.33  \\ 
\multicolumn{1}{|l|}{$f_{\rm Edd,2},\log(M_{\rm gal,2})$} & \multicolumn{1}{l|}{5.332E-21}       & \multicolumn{1}{l|}{2.049E-20}       & \multicolumn{1}{l|}{18.60}              & 1.71   & 243.60             & 13.39  & 37.60              & 2.84  & 553.30             & 4.32  \\ \hline
\multicolumn{5}{|l|}{1D,$z=2$:  $N_d$=$N_c$=1310, $N_a$=32001}                                                                                                                             &                    &        &                    &       &                    &       \\ \hline
\multicolumn{1}{|l|}{$f_{\rm Edd,1},\log(M_{\rm BH,1})$}  & \multicolumn{1}{l|}{1.341E-15}       & \multicolumn{1}{l|}{1.517E-73}       & \multicolumn{1}{l|}{12.40}              & 1.51   & 147.50             & 15.01  & 6.800              & 0.422 & 60.50              & 1.08  \\ 
\multicolumn{1}{|l|}{$f_{\rm Edd,2},\log(M_{\rm BH,2})$}  & \multicolumn{1}{l|}{2.467E-12}       & \multicolumn{1}{l|}{2.360E-10}       & \multicolumn{1}{l|}{21.80}              & 2.70   & 194.10             & 23.76  & 44.900             & 1.449 & 570.50             & 5.08 \\ 
\multicolumn{1}{|l|}{$f_{\rm Edd,1},\log(M_{\rm gal,1})$} & \multicolumn{1}{l|}{1.983E-10}       & \multicolumn{1}{l|}{4.270E-75}       & \multicolumn{1}{l|}{14.30}              & 1.89   & 192.50             & 16.55  & 5.500              & 0.527 & 60.50              & 1.27  \\ 
\multicolumn{1}{|l|}{$f_{\rm Edd,2},\log(M_{\rm gal,2})$} & \multicolumn{1}{l|}{4.157E-12}       & \multicolumn{1}{l|}{2.190E-13}       & \multicolumn{1}{l|}{18.80}              & 2.20   & 193.60             & 24.37  & 30.700             & 2.791 & 412.10             & 3.25  \\ \hline
\multicolumn{5}{|l|}{2D,$z=1$:  $N_d$=$N_c$=626, $N_a$=23252}                                                                                                                              &                    &        &                    &       &                    &       \\ \hline
\multicolumn{1}{|l|}{$f_{\rm Edd,1},\log(M_{\rm BH,1})$}  & \multicolumn{1}{l|}{7.515E-06}       & \multicolumn{1}{l|}{8.570E-32}       & \multicolumn{1}{l|}{15.10}              & 2.08   & 178.70             & 26.12  & 6.500              & 0.527 & 61.80              & 0.92  \\ 
\multicolumn{1}{|l|}{$f_{\rm Edd,2},\log(M_{\rm BH,2})$}  & \multicolumn{1}{l|}{5.641E-06}       & \multicolumn{1}{l|}{1.717E-08}       & \multicolumn{1}{l|}{18.80}              & 1.99   & 174.60             & 25.26  & 29.600             & 2.171 & 331.90              & 3.90  \\ 
\multicolumn{1}{|l|}{$f_{\rm Edd,1},\log(M_{\rm gal,1})$} & \multicolumn{1}{l|}{1.553E-05}       & \multicolumn{1}{l|}{2.300E-36}       & \multicolumn{1}{l|}{19.80}              & 2.39   & 217.00             & 18.76  & 6.700              & 0.483 & 68.30             & 0.95 \\ 
\multicolumn{1}{|l|}{$f_{\rm Edd,2},\log(M_{\rm gal,2})$} & \multicolumn{1}{l|}{2.758E-04}       & \multicolumn{1}{l|}{1.226E-10}       & \multicolumn{1}{l|}{23.90}              & 2.64   & 251.20             & 39.49  & 27.300             & 1.059 & 307.30             & 2.83  \\ \hline
\end{tabular}
\caption{Results of one- and two-dimensional Kolmogorov-Smirnov tests on distributions of Eddington ratios, galaxy and MBH masses comparing the sample of dual AGN  with the control sample and with the full AGN population  at $z=1,2,3$. $N_d$ and $N_c$ are the (identical) sizes of the dual and control sample, $N_a$  is the size of the full AGN population sample. $\mathcal{P}$,c is the probability that the dual and control sample are extracted from the same parent population, $\mathcal{P}$,a is the same for the dual and the full AGN population samples. $N_{d,c}$(1$\sigma$) is the minimal (identical) size of subsamples extracted from $N_d$ and $N_c$ needed to show at more than 1$\sigma$ level that the two distributions are not extracted from the same population, $N_{d,c}$(3$\sigma$) is the analog for the 3$\sigma$ and $N_{d,c}(0.99)$ at the 0.99 per cent level. $N_{d,a}$ follows the same convention, but for subsamples extracted from $N_d$ and $N_a$.}
\label{tab:KS}
\end{table*}

\subsection{Quantitative analysis on the distinctive properties of dual AGN}

In the previous two sections we have described -- qualitatively -- the properties of dual AGN and compared them -- qualitatively -- to the general AGN population. However, selecting the brighter (fainter) of a pair of AGN, even if they are far apart and thus unrelated, will bias the distribution to higher (lower) luminosity. The most luminous AGN in a random pair should have higher MBH mass and/or higher Eddington ratio and viceversa for the least luminous AGN. In this section we assess whether results are driven by this bias and provide a quantitative comparison with control samples and with the general AGN population. 
 
\begin{figure*}
	\includegraphics[width=\columnwidth]{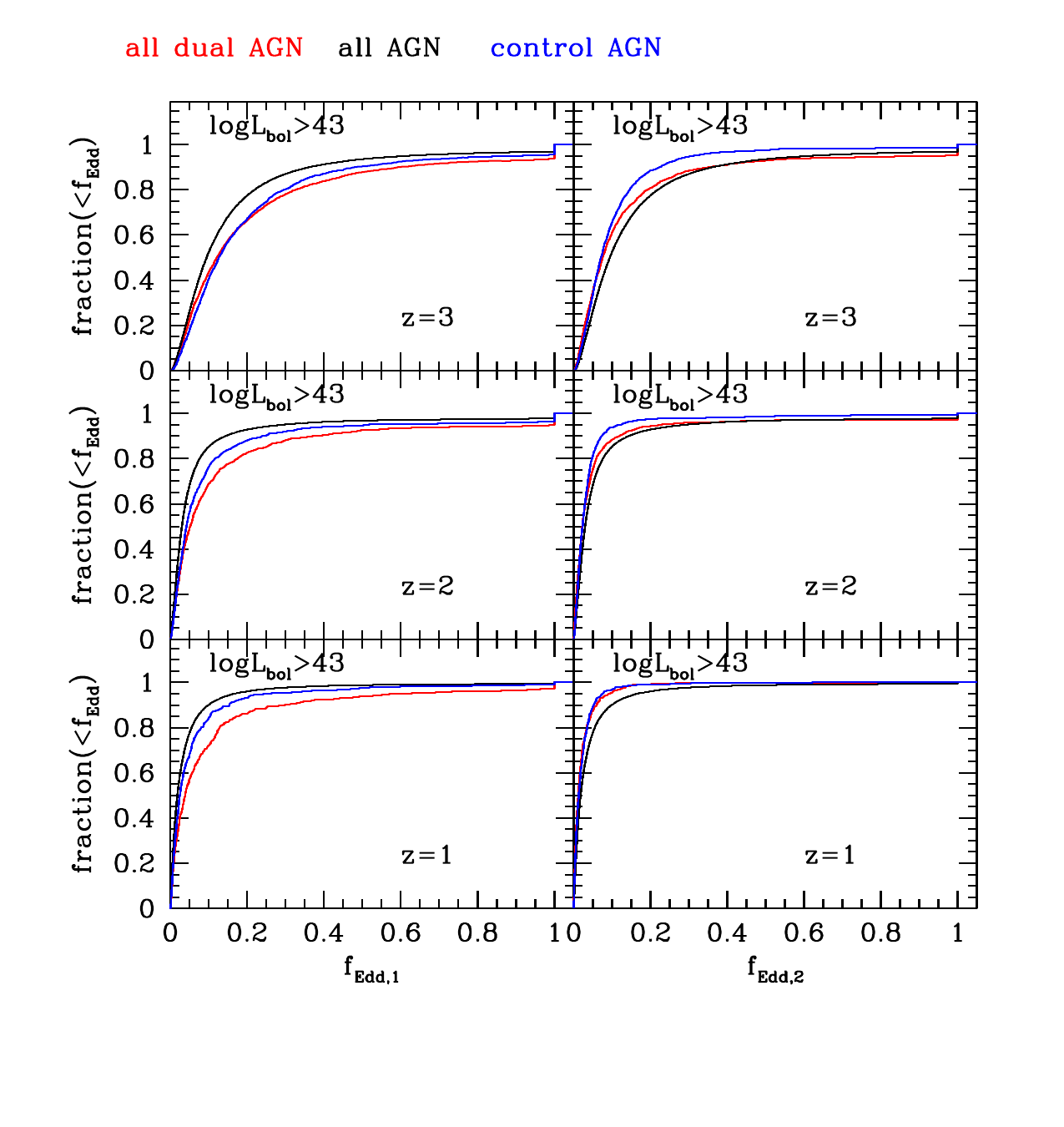}
	\includegraphics[width=\columnwidth]{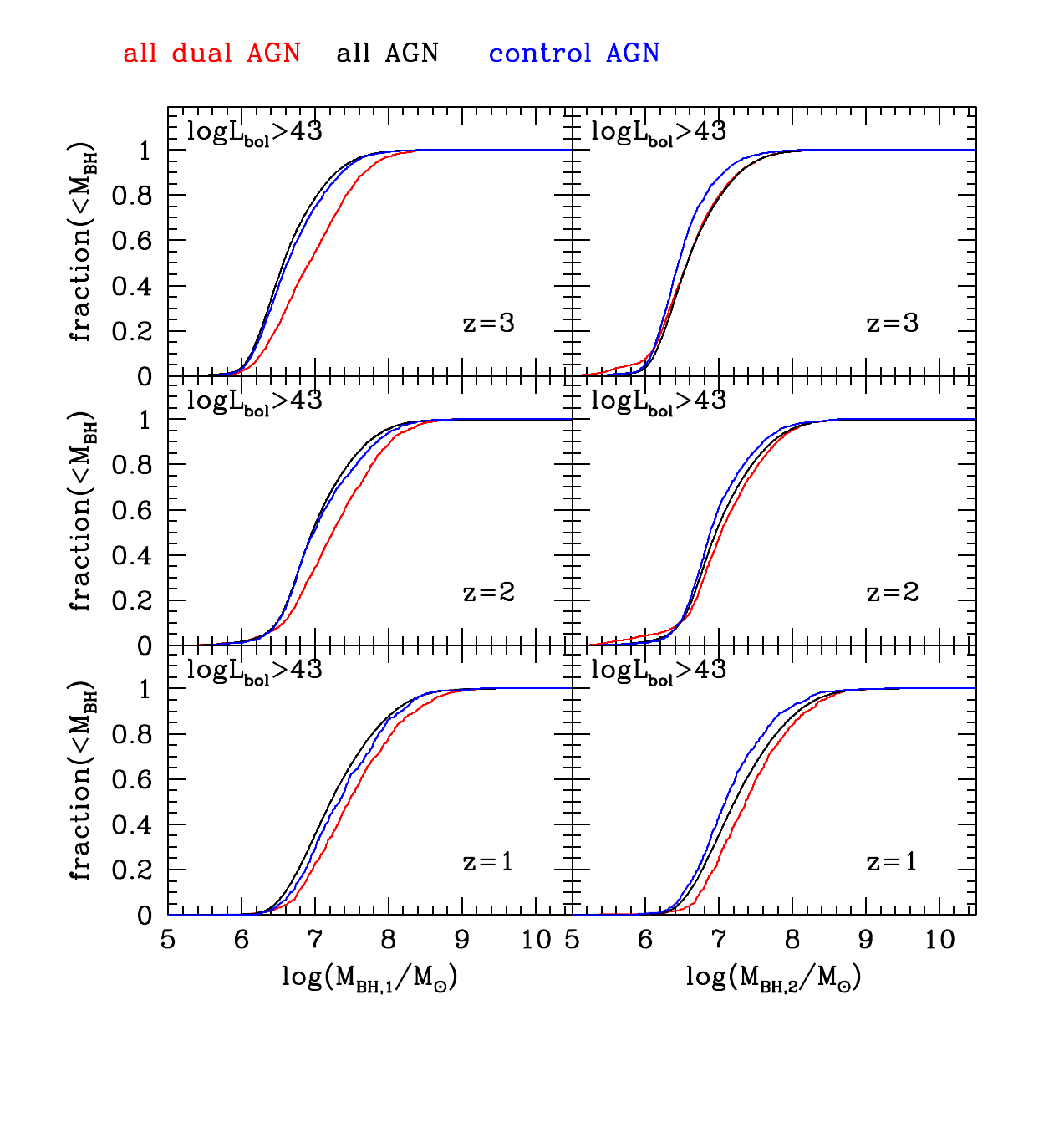}
	\includegraphics[width=\columnwidth]{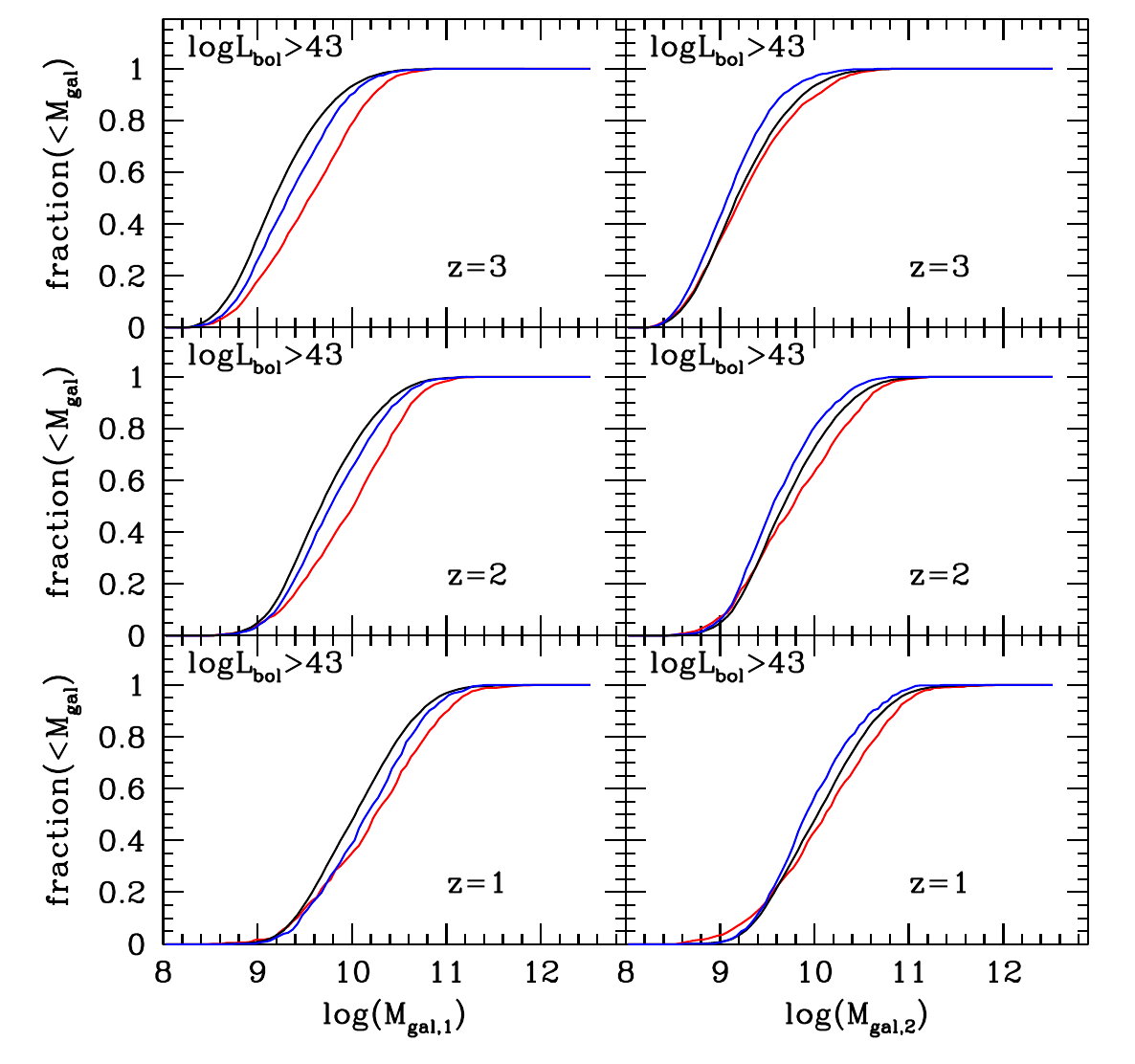}
	\includegraphics[width=\columnwidth]{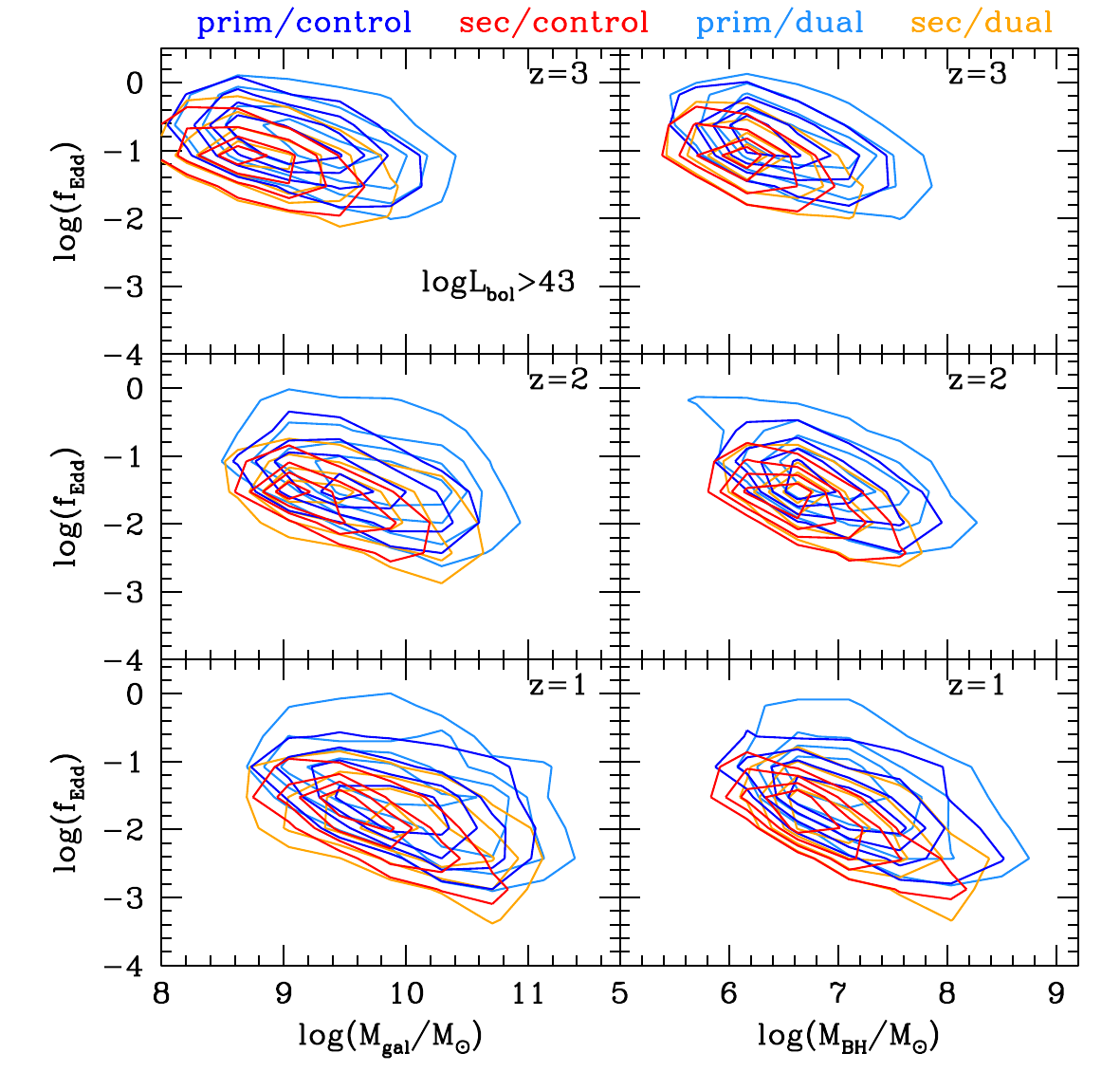}
    \caption{Comparison of AGN properties for dual AGN, a control sample and the full AGN population. The control sample selects two random AGN from the full AGN population to create artificial pairs that are not physically related.}
    \label{fig:control}
\end{figure*}

We have constructed ten control samples by selecting random pairs of AGN from the entire box, with the same luminosity cut and number of objects as in the dual AGN samples at the same redshift. In the control and general population samples we cannot divide them into ``different'' and ``same'' galaxy. In the following analysis therefore the dual sample includes both duals hosted in one and two galaxies. We note however that dual AGN in different galaxies largely outnumber dual AGN in one galaxy (at least using our criteria).  We  perform the analyses on the main sample ($\log(L_{\rm bol})>43$) without imposing additional cuts. The distributions of MBH and galaxy masses and of the Eddington ratio for dual AGN, the full AGN population and one of the control samples are shown in Fig.~\ref{fig:control}. 

In Table~\ref{tab:stats} we report the mean and standard deviation of MBH and galaxy masses (in log space) and of the Eddington ratio. The first item to note is that in all cases the standard deviation is larger than the difference between the samples. The mean values confirm the qualitative trends described in the previous sections, but the standard deviation shows that they all have a low statistical significance.  The comparison with the control sample shows similar results, except that the difference in primary masses  (MBH and galaxy) and Eddington ratios are smaller, as expected. For secondaries, instead, the trends for the mean values in masses are reversed in the dual and control samples:  the control sample behaves as expected, in the sense that MBHs/galaxies in the control are less massive than in the general population, while in the dual sample MBH/galaxies are more massive than in the general population. This suggests that the dual sample is not subject to the selection bias described above, but also in this case the large standard deviation limits the statistical significance of the results. 

We have also performed Kolmogorov-Smirnov (KS) tests in one and two dimensions\footnote{For the 2D KS test, the results are less trustworthy when the correlation coefficient of the two distributions differ \citep{1987MNRAS.225..155F,1992nrfa.book.....P}. This is the case for some of the 2D distributions, in particular $f_{\rm Edd,2}$,$\log(M_{\rm gal,2})$ at $z=3$ (where a large number of high $f_{\rm Edd,2}$ systems at high $\log(M_{\rm gal,2})$ exist), and to a lesser extent at $z=2$ and $z=1$. In this case the uncertainty in the significance level is of order 5 per cent.} to assess the differences (or not) between the distributions. The results are reported in Table~\ref{tab:KS}. In the KS tests, we first calculate the probability that the two samples are drawn from the same parent distribution.  Only $\log(M_{\rm gal,1})$ and $f_{\rm Edd,2}$ for the dual/control samples at $z=1$ and $f_{\rm Edd,2}$ for the dual/control samples at $z=2$ have a probability higher than 0.003 (i.e. a significance level lower than $>3-\sigma$) of originating from the same parent distribution, while $\log(M_{\rm gal,1})$  for the dual/control sample at $z=1$ is borderline.  

We then calculate the minimal sample size needed to prove that the distributions differ at 1- and 3-$\sigma$ level, or at 0.99 per cent level in the case of 2D distributions \citep[the maximum confidence level provided by][]{1987MNRAS.225..155F} . We calculated this by creating 1000 realizations for each subsample size and defining the minimal sample as that for which the mean probability drops below the required level. We have checked that convergence is reached for 1000 realizations. The standard deviation is obtained over 10 different control samples and 10 different random number draws for the full sample.  We note that there can be apparent inconsistencies between the full analysis and the analysis on subsamples because in the full analysis the number of objects in the general AGN population sample is much larger than the number of objects in the dual sample, while in the subsample analysis we force the size of both samples to be the same for simplicity. 

One could interpret the results by considering that the smaller the sample needed to distinguish the distributions, the more they differ. It thus appears to be relatively easy to distinguish the properties of primary AGN in duals from the general population, while for secondary AGN this requires a much larger sample (sometimes larger than the simulated sample), implying that the properties of primary AGN in duals are more dissimilar than those of secondary AGN with respect to those of the general AGN population. When comparing dual and control 2D distributions there is not much difference between primaries and secondaries in terms of sample size needed to distinguish the two distributions, with some exceptions (see Table~\ref{tab:KS}). In summary, primary AGN in duals differ more from the general AGN population than secondary AGN do. Secondary AGN in duals differ (relatively) more from the control sample than from the all sample: this is driven by their MBH/galaxy masses being larger than for the AGN in the control sample, hinting that dual AGN indeed prefer massive galaxies and MBHs. 

\section{Multiple AGN and their environment}

\begin{figure}
	\includegraphics[width=\columnwidth]{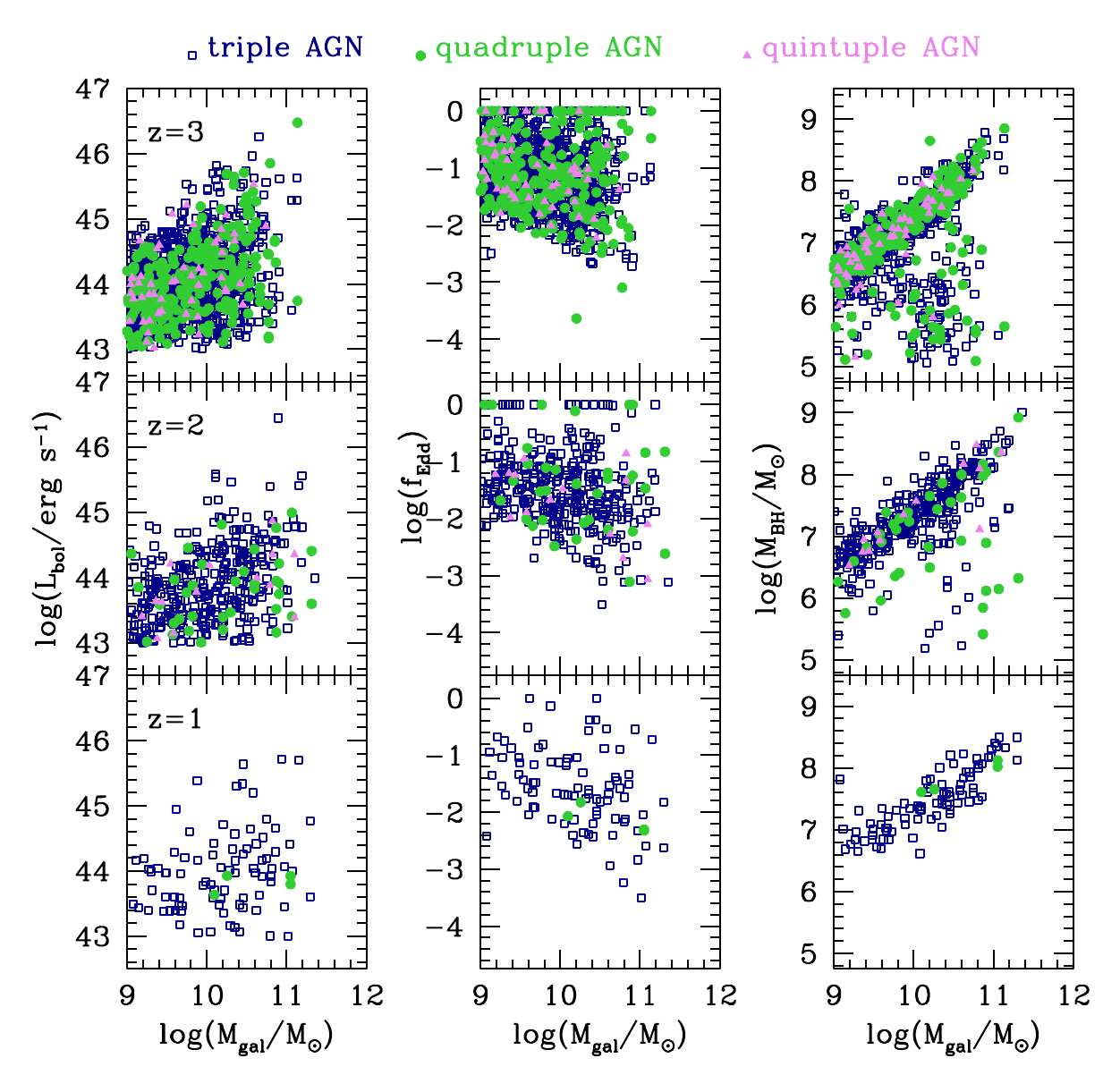}
    \caption{Properties of triple, quadruple and quintuple AGN where each AGN has  $\log(L_{\rm bol})>43$. Bolometric luminosity, Eddington ratio and MBH mass are shown as a function of galaxy stellar mass.}
    \label{fig:trip_penta}
\end{figure}

\begin{figure}
	\includegraphics[width=\columnwidth]{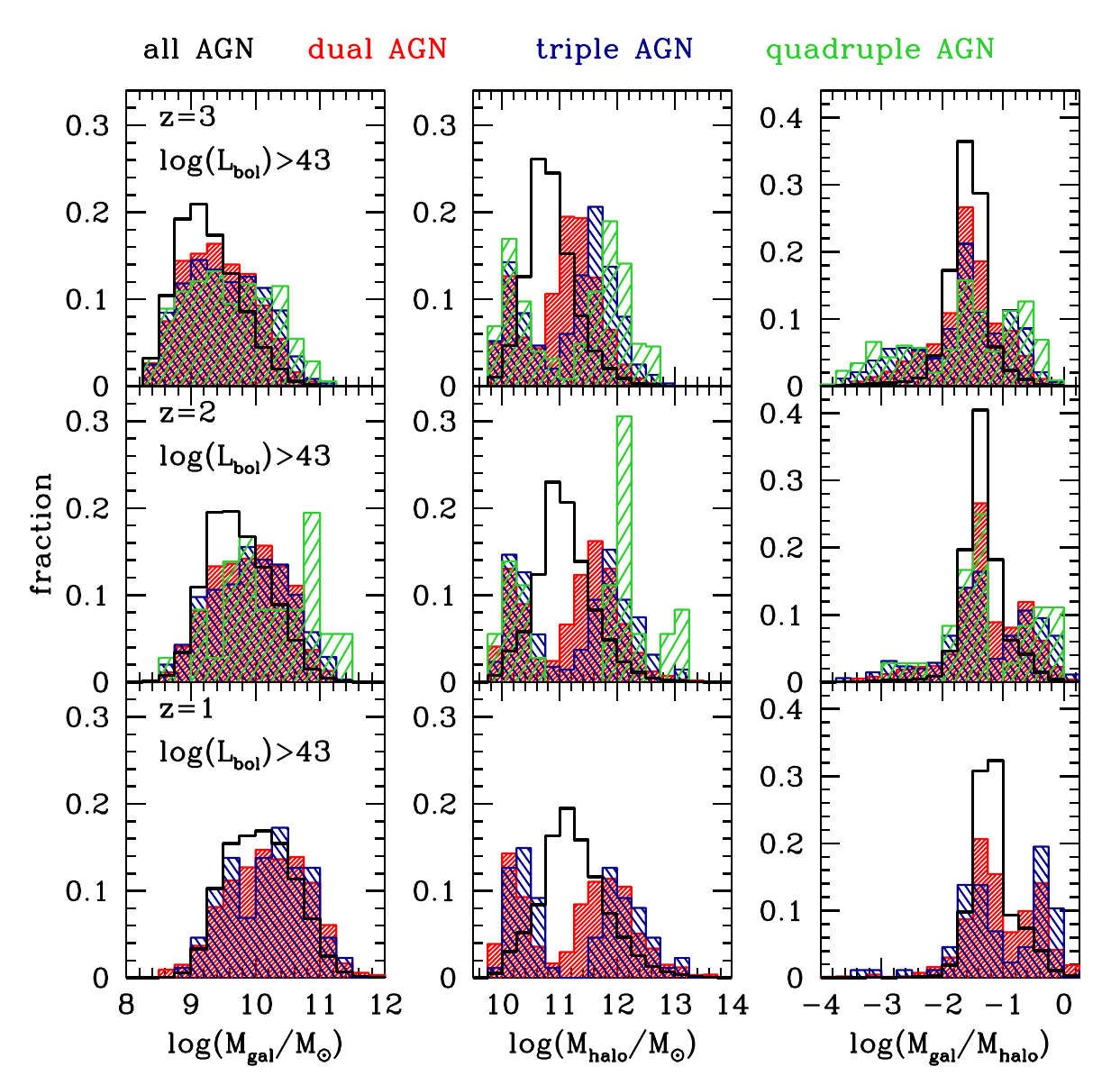}
    \caption{For each AGN in a multiple system their galaxy and halo hosts are shown and compared to galaxy and halo hosts of the whole AGN population. The bimodal distribution of halo masses is caused by all halos with mass $<10^{11} \msun$ being sub-halos of the larger halo hosting one of the other AGN in the multiplet. The galaxy/halo mass ratios present significant tails both towards very high or low mass ratios because of sub-halo mass loss: for galaxies still surrounded by a sub-halo the mass ratio increases, while for galaxies whose sub-halo has been completely disrupted the mass ratio becomes very small (in this case we associate the galaxy to the main halo, as a subgalaxy). In general, the higher the multiplicity, the higher the mass of the (main) halo.}
    \label{fig:gal_halos_multi}
\end{figure}

We show the basic properties of multiple AGN systems in Fig.~\ref{fig:trip_penta}.  Comparison with Figures~\ref{fig:mbh_mgal} and~\ref{fig:fedd_mass} shows that multiple systems have properties similar to dual AGN. Generally, the MBHs significantly below the mass expected from the relation with the host galaxy mass are hosted in the same galaxy with a more massive MBH that sits on the relation, and are wandering MBHs or MBHs on the way to coalescence \citep[see][for similar results in a semi-analytical model]{2009MNRAS.400.1911V}. 

Most multiple systems are expected to be hosted in separate galaxies \citep{2020MNRAS.492.5620B}. Using our standard threshold of $\log(L_{\rm bol})>43$, we find that the fraction of triple systems in separate galaxies is 80 percent at $z=3$ to 70 percent at $z=1$, whereas below $z=1$ there are too few galaxies for statistics. The rest are hosted two in a galaxy and the third in a separate galaxy. The fraction of triples hosted in one single galaxy is less than 3 per cent. The picture for quadruple systems is similar: the majority is hosted in four separate galaxies, although by $z\sim 1.5$ quadruple AGN hosted in three separate galaxies become more common. Below $z=1.5$ there are between 0 and 1 quadruple systems. 

A natural question is whether multiple AGN are more common in galaxy groups, i.e., large halos hosting several galaxies \citep{2020MNRAS.492.5620B}. In Fig.~\ref{fig:gal_halos_multi} we compare how the galaxy and halo properties of the AGN population depend on multiplicity. The mass of the host galaxies simply extends to larger masses the higher the multiplicity (left panel), whereas the halo mass (middle panel), while also extending to larger masses the higher the multiplicity, has a bi-modal behaviour for multiple AGN. The halo masses of the whole population -- which is dominated by single AGN -- peaks at about $10^{11} \msun$, while the halo masses of multiple AGN avoid masses around $10^{11} \msun$, and are either smaller or, for the most part, larger. The halos in the low mass peak proved indeed to be all sub-halos, while the halos in the high mass peak are predominantly main halos. Using the ratio of galaxy to halo mass (right panel) as a further indication, it becomes apparent that there are three types of galaxies in groups that participate in the multiple AGN population. First, the central galaxy of the group, which is part, for instance, of a triple AGN in about 70 per cent of cases. Then, galaxies that are embedded in sub-halos, which in many cases have started loosing mass because of interactions in the groups' potential. These systems have unusually high galaxy-to-halo mass ratio. Finally, galaxies that have lost completely their sub-halo mass and are now associated directly to the halo of the group, as sub-galaxies. These systems have very low galaxy-to-halo mass ratio. 

In summary, dual and multiple AGN are linked to at least one massive halo, and the higher the multiplicity the higher the mass of the main halo. When looking for multiple systems, targeting galaxy groups is therefore expected to give a higher success rate than targeting blank fields. 

\section{Dual AGN as tracers of galaxy and MBH mergers}

\begin{figure}
	\includegraphics[width=\columnwidth]{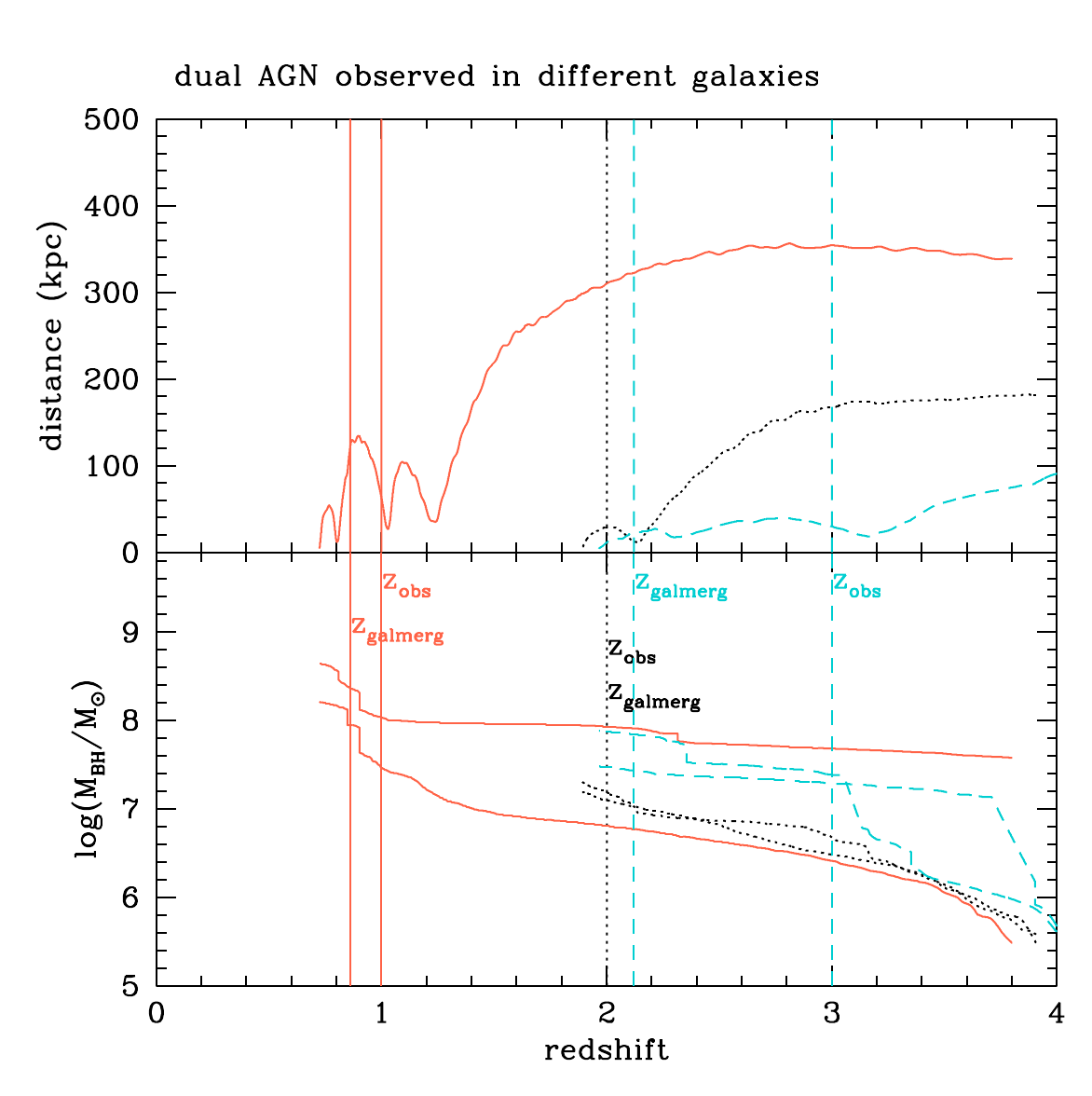}
	\includegraphics[width=\columnwidth]{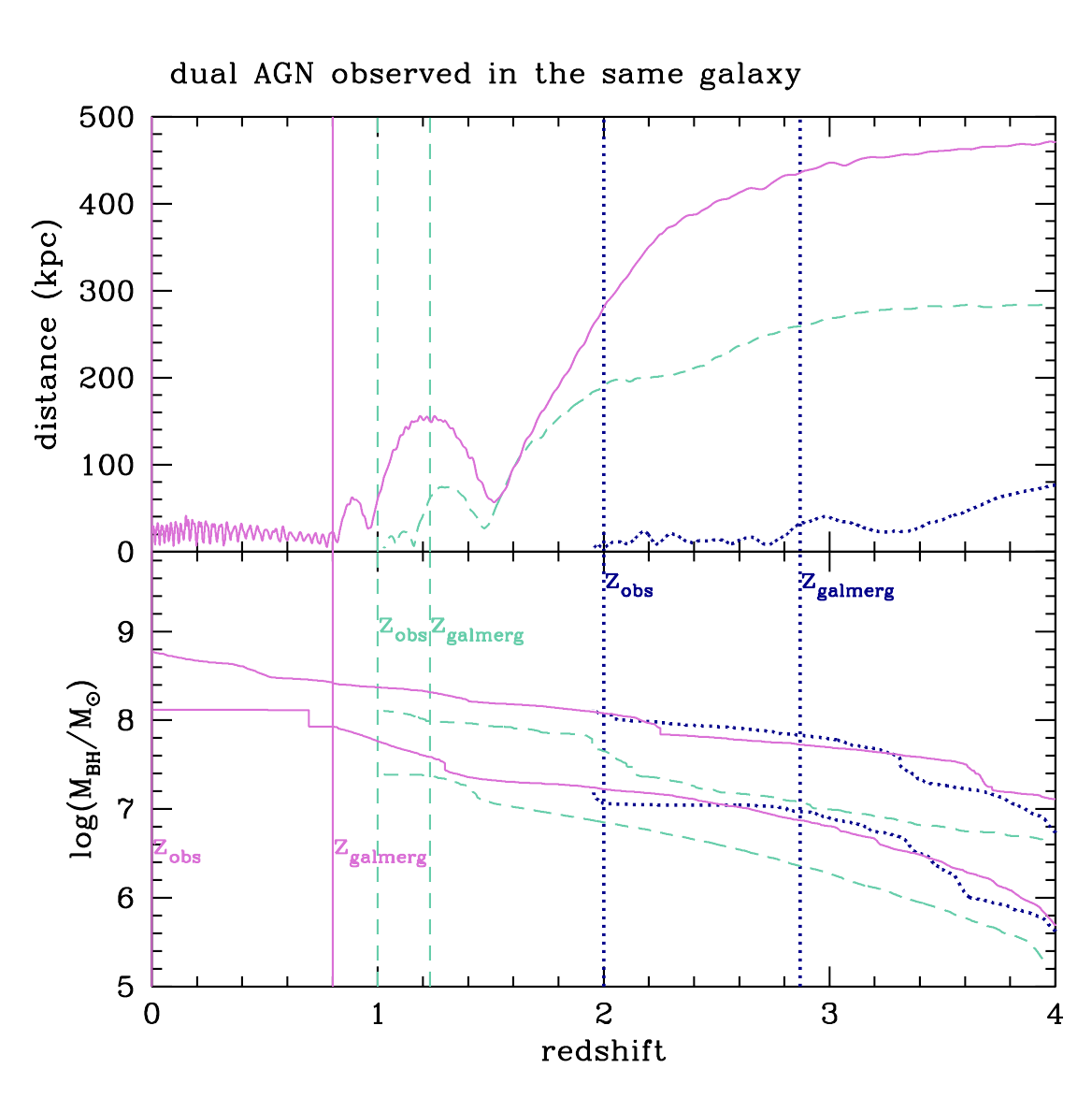}
    \caption{Examples of the evolution of dual AGN, in different (top) or in the same (bottom) galaxy at the time of observation. Distance between the two MBHs and MBH masses are shown as a function of redshift, with different colors and line styles for different dual AGN. In each panel vertical lines trace the redshift of observation ($z_{\rm obs}$, i.e., the redshift at which we select the dual AGN) and the redshift of the galaxy merger linked to origin of the dual AGN ($z_{\rm galmerg}$). If the curves for a pair terminate, the MBHs have merged at that redshift.}
    \label{fig:examples}
\end{figure}

\begin{figure}
	\includegraphics[width=\columnwidth]{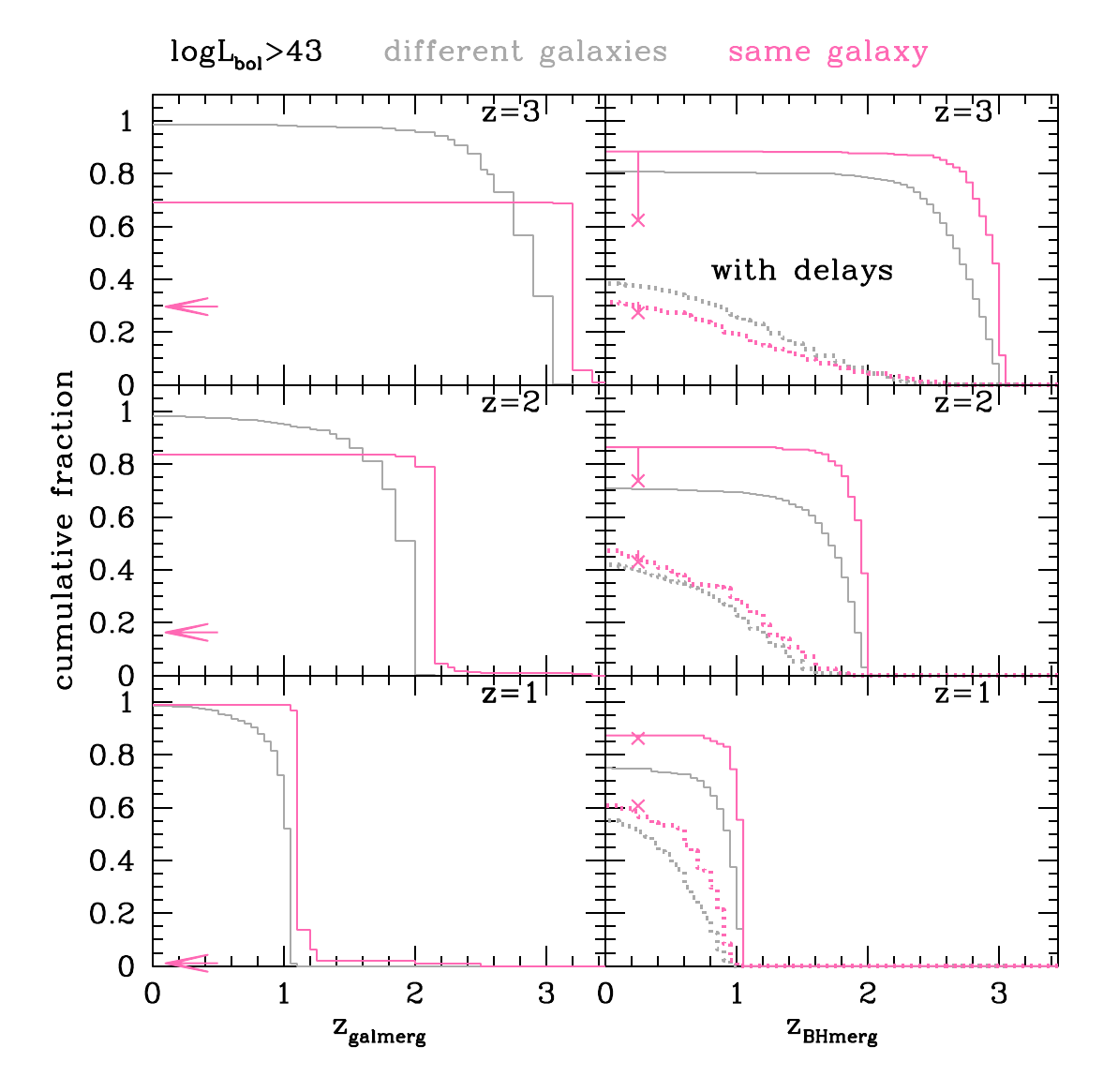}
    \caption{For dual AGN  observed at a given redshift (labelled in each panel) the figure shows the cumulative distribution of the redshifts of the related galaxy (left) and MBH (right, solid without accounting for dynamical delays in postprocessing; dotted accounting for delays) mergers. Grey histograms refer to dual AGN hosted in different galaxies and pink ones to dual AGN hosted in the same galaxy. Most dual AGN at low redshift can be related to a galaxy merger, although the time between the galaxy merger and the observation of the dual can vary. At high redshift a significant fraction of dual AGN hosted in the same galaxy cannot be traced back to a galaxy merger (shown with an arrow, see Section~\ref{sec:LinkingDualAGNToGalaxyMergers}), and a few dual AGN hosted ``in the same galaxy'' have $z_{\rm galmerg}<z$ (see Section~\ref{sec:LinkingDualAGNToGalaxyMergers}). In the left panel the crosses mark how the fraction of MBH mergers changes if we exclude the dual AGN that cannot be traced back to a galaxy merger.  Not all dual AGN give rise to a MBH merger, especially if we account for sub-kpc delays. If we limit the analysis to dual AGN hosted in the same galaxy the probability of MBH mergers increases to $\sim$ 80 per cent when not accounting for dynamical delays, when delays are included the difference for duals in one or two galaxies decreases.  
    }
    \label{fig:zdual_zBHmerge_zgalmerge_diffgal}
\end{figure}

\begin{figure}
	\includegraphics[width=\columnwidth]{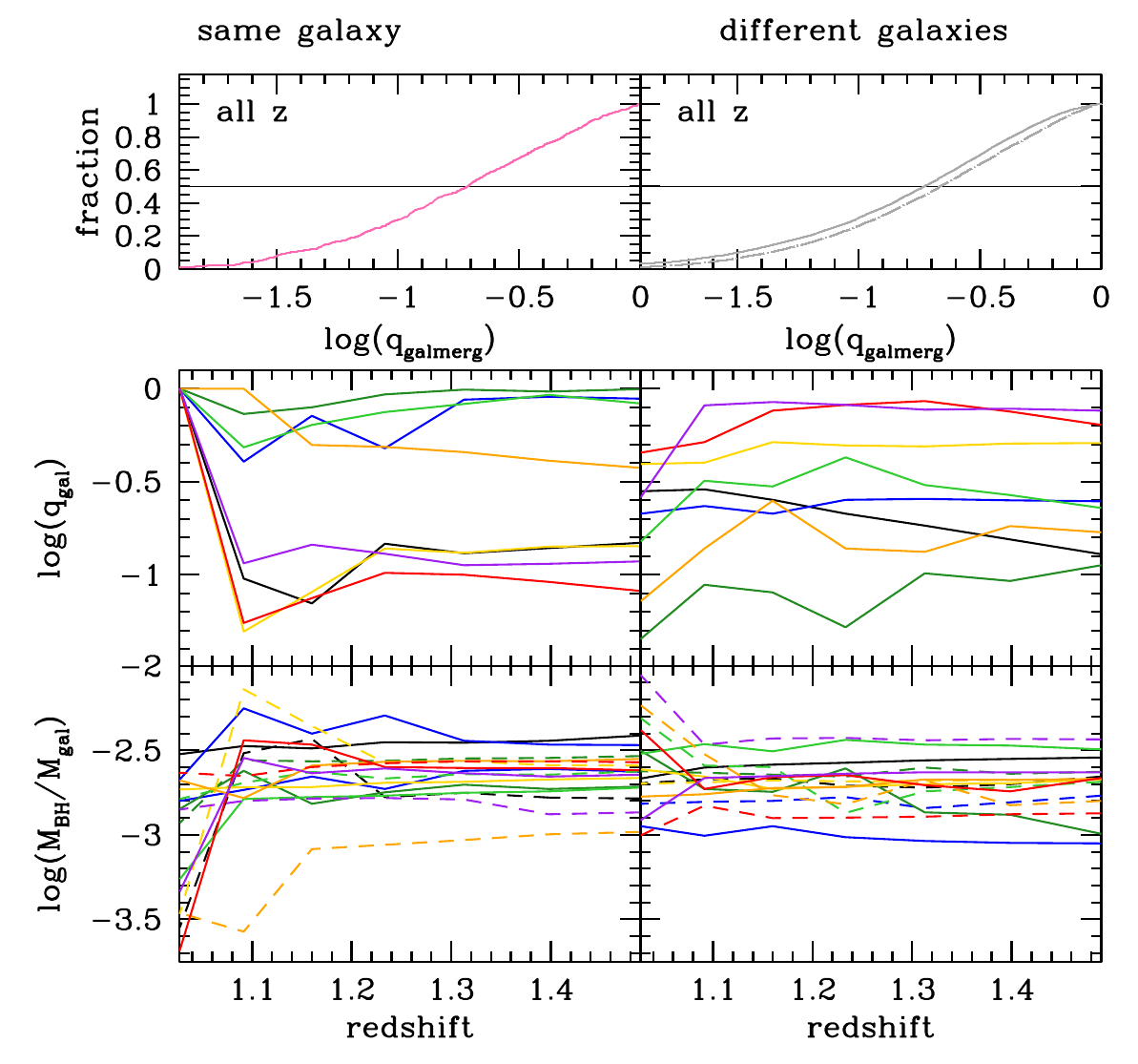}
    \caption{Top panels: cumulative distribution of the mass ratio of galaxy mergers linked to dual AGN, for all analyzed redshifts. For duals observed in different galaxies we show also, with a dot-dashed curve, the mass ratio at the time of observation. The horizontal line marks 50 per cent. Middle panels: for a representative set of duals observed at $z=1$ the evolution with redshift of the galaxy mass ratio is shown. In many cases the mass ratio decreases approaching the merger, a combination of the larger galaxy gaining mass via star formation and the smaller galaxy loosing mass via stripping. Bottom panels: for the same duals shown in the middle panels, we show tracks of the evolution of MBH-to-galaxy mass ratio (the MBH powering the primary AGN with solid curves, the secondary's MBH with dashed curves). Stripping leads to ``overmassive'' MBHs, while when the galaxies merge the MBH previously hosted in the smaller galaxy appears as ``undermassive'' with respect to the merger remnant.}
    \label{fig:mgal_mbh_ratio_histories_distr}
\end{figure}

Dual AGN have been proposed to be used as tracers of galaxy mergers \citep{2009ApJ...698..956C}, and there has been discussion on whether they can also be predictors for MBH mergers. In this section we analyze the link between dual AGN, observed at a given time, and whether they can be connected to a galaxy or MBH merger.  

For dual AGN hosted in different galaxies we search forward in time for whether the galaxies will merge. For dual AGN hosted in the same galaxy we check if their origin can be traced back to a galaxy merger  (see \S\ref{sec:LinkingDualAGNToGalaxyMergers}). In analogy with galaxies, we search for a MBH merger that involved the two MBHs powering a dual AGN. We consider both an optimistic approach, where MBHs are considered merged when they coalesce in the simulation (``numerical merger''), which happens at a distance of $4\Delta x=4$~kpc, and a conservative approach, where the time of the merger is calculated in post-processing adding dynamical delays \citep[``delayed merger'', see][for details]{2020MNRAS.498.2219V}. 

We are selecting dual AGN that are observable, during their evolution, at $z_{\rm obs} =0, \,0.5, \,1, \,1.5, \, 2, \, 2.5, \, 3.$ In this sense these are arbitrary times and they are not related to a specific phase in the evolution of the system: in observations one is only able to glance at a snapshot of the full evolution.  In the simulation, however, we can follow the evolution both before and after. The  population includes rapidly evolving systems, slower ones, and even ineffective MBH mergers giving rise to wandering MBHs. Figure~\ref{fig:examples} shows the variety of situations that can be encountered. 

\subsection{Dual AGN and galaxy mergers}

\begin{figure}
	\includegraphics[width=\columnwidth]{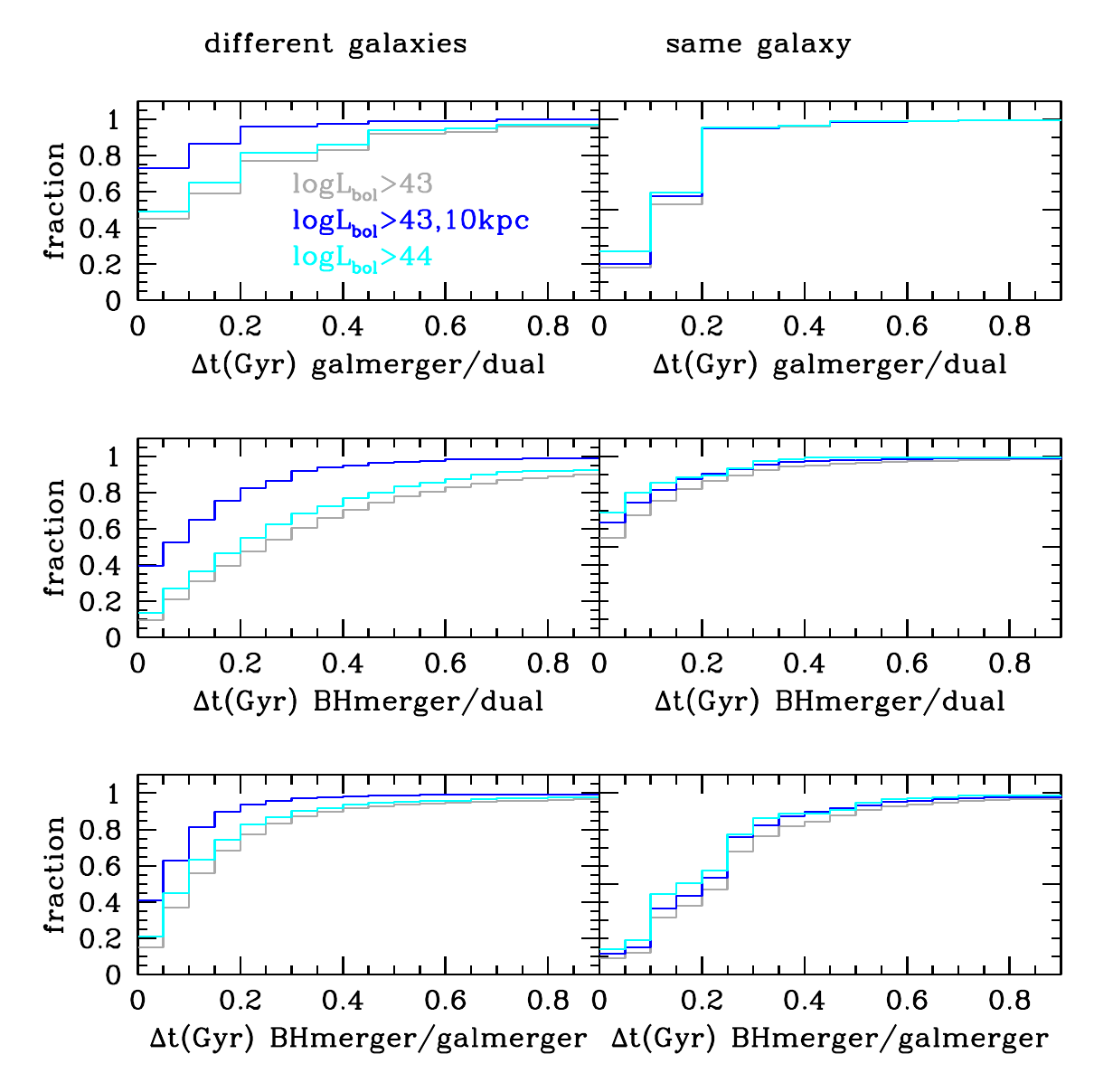}
    \caption{Time elapsed between galaxy merger and dual AGN observation (top), between MBH merger and dual AGN observation (middle), between galaxy and MBH merger (bottom), for all dual AGN that can be matched to both a galaxy and a MBH merger. Duals with small separations and/or higher luminosities provide a better selection for rapidly evolving mergers (both galaxies and MBHs), especially for duals hosted in different galaxies.
    }
    \label{fig:deltat}
\end{figure}

In the left panel of Fig.~\ref{fig:zdual_zBHmerge_zgalmerge_diffgal} we summarize how faithfully dual AGN can trace galaxy mergers. For dual AGN hosted in different galaxies, eventually the two galaxies will merge, and in 90 per cent of cases the galaxy merger occurs within 0.6-1~Gyr from the time of observation of the dual AGN, but for a minority of cases the delay can be up to 5~Gyr. 

Our analysis is in agreement with previous analyses of cosmological simulations \citep{2016MNRAS.458.1013S,2019MNRAS.483.2712R} that dual AGN are related to mergers with a substantial mass ratio. We find that the typical mass ratio is 0.2 (Fig.~\ref{fig:mgal_mbh_ratio_histories_distr}, top panels). For duals in different galaxies the mass ratio at the time of observation is somewhat larger than the mass ratio at the time of the merger. This is because in the intervening time the smaller galaxy looses mass, by stripping, while the larger galaxy gains mass, via star formation. To illustrate this, the evolution of the mass ratio of the galaxies for a representative set of dual AGN observed at $z=1$ is shown in the middle panels of Fig.~\ref{fig:mgal_mbh_ratio_histories_distr}. The bottom panels show how the ratio of MBH to galaxy mass evolves over the same redshift span. Significant changes occur when a galaxy looses mass (and the MBH becomes ``overmassive'') or at the time of the galaxy merger, when the smaller MBH is associated to a much larger galaxy, the merger remnant (and the MBH becomes ``undermassive'').  The typical mass ratio of the MBHs is 0.2 for duals in different galaxies and 0.3 for duals in one galaxy. Since the Eddington ratio of the MBH in the smaller galaxy is generally higher than the Eddington ratio of the MBH in the larger galaxy, the mass ratio has a  tendency to increase with time \citep[in agreement with][]{2015MNRAS.447.2123C}, contrary to the galaxy mass ratio.

A fraction of dual AGN observed in the same galaxy cannot be traced to a previous galaxy merger. This is related to purely numerical reasons (at least in this analysis). There are two situations that lead to this events. The first is a MBH that forms in a dense gas cloud in the outskirts of a galaxy or in a filament. The exclusion radius for MBH formation is 50~comoving~kpc, which is less than our distance threshold of 30~kpc for $z>0.67$. Such new MBH, forming in a dense gas cloud, would start accreting immediately and would be selected as a dual AGN if the galaxy to which is matched already contains an AGN. MBH formation is stopped at $z=1.5$, therefore these systems disappear afterwards. The second case for a dual AGN is when a MBH forms at very high redshift in a cloud unassociated with a galaxy (because it it too small to meet the criteria in the halo finder), and travels without being captured by a galaxy, until at some point the MBH gets close enough to a galaxy to be associated to it. If the galaxy contains another AGN, the dual AGN algorithm picks this, by construction, and associates the two MBHs to the same galaxy, but there is no merger event delivering the second MBH: no galaxy merger was ever involved (see also Section \ref{sec:LinkingDualAGNToGalaxyMergers}). Increasing the luminosity threshold for dual AGN selection decreases the occurrence of such systems, since most of them include a low-mass MBH; decreasing the distance between duals for selection also helps, since such MBHs are typically in the outskirts of galaxies (See the Appendix). 

When investigating the time difference between dual observation and galaxy merger (Figure~\ref{fig:deltat}), as expected, for duals in different galaxies a smaller separation hints that the galaxy will follow shortly. Less obviously, this is also the case for duals hosted in the same galaxy, albeit to a lesser extent.  Although one would expect a longer time after a galaxy merger for MBHs to be separated by a smaller distance, the MBHs that reach small separation are the subset that have an efficient orbital decay.  Higher luminosity duals also trace mergers that are closer in time, because higher masses and accretion rates lead to faster dynamical evolution.

\subsection{Dual AGN and MBH mergers}\label{sec:AGNMBH}

 \begin{figure}
 	\includegraphics[width=\columnwidth]{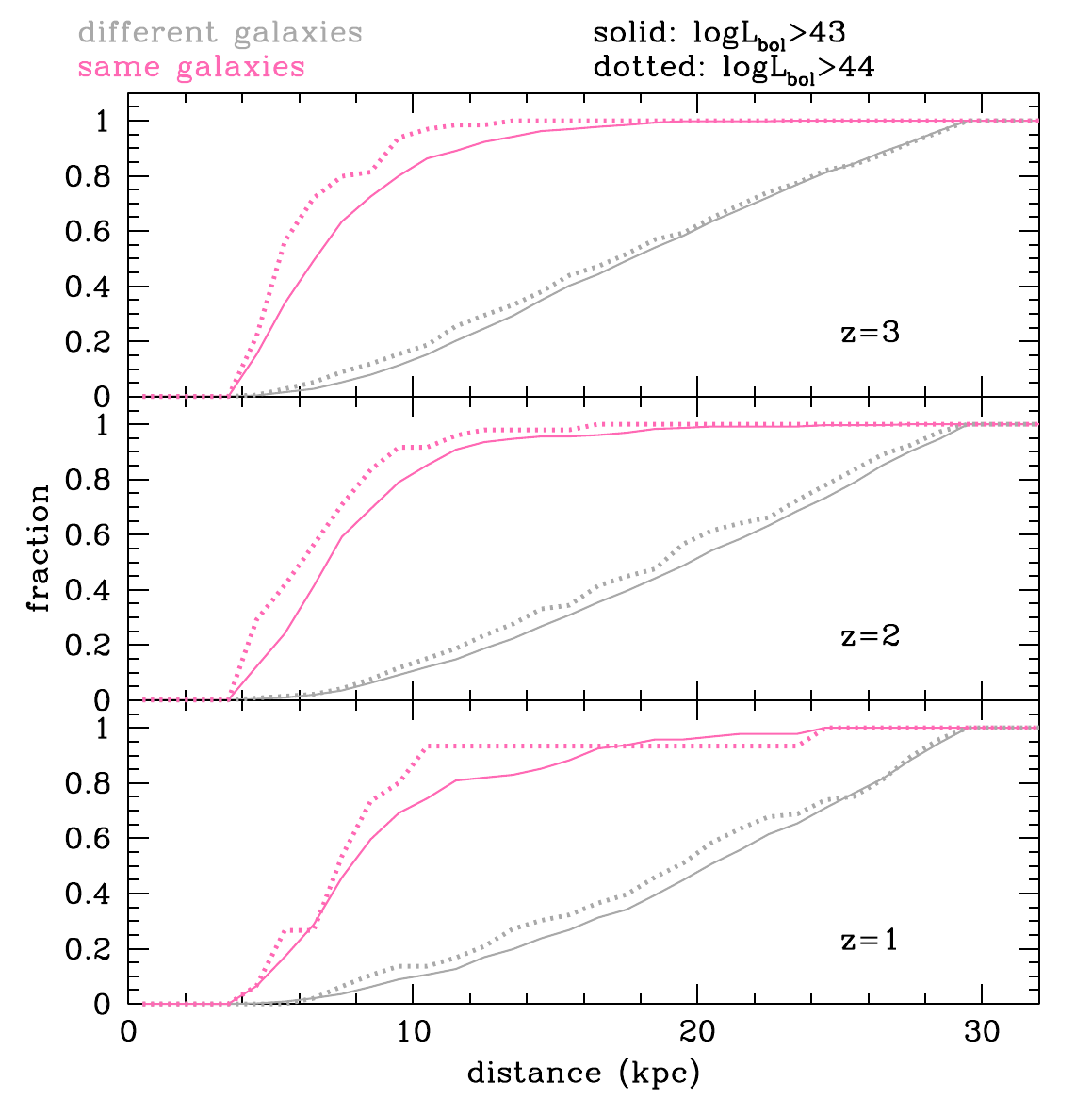}
     \caption{Cumulative distribution of dual AGN distances, for two luminosity cuts, and for duals hosted in different galaxies or in one single galaxy. 
     Dual AGN with small separations are much more likely (but not necessarily) hosted in the same galaxy. High luminosity duals in the same galaxy are preferentially found at small (<10~kpc) separations: high MBH masses and high accretion rates -- thus high gas and presumably stellar densities --  favour effective orbital decay.}
     \label{fig:dist_same_diff_lum}
\end{figure}

The capability of dual AGN to be considered precursors of MBH mergers is investigated in the right panel of Fig.~\ref{fig:zdual_zBHmerge_zgalmerge_diffgal}. 
The failure to connect all dual AGN to a MBH merger is not surprising: not all galaxy mergers end in a MBH merger because MBH dynamics can be inefficient on both large \citep{1994MNRAS.271..317G,2017ApJ...840...31D,2018MNRAS.475.4967T,2019MNRAS.486..101P,2020ApJ...896..113L,2020MNRAS.498.3601B, 2021arXiv210307486B} and small \citep{BBR1980,Milosavljevic2001,2019ApJ...871...84M} scales.  

If\footnote{In this analysis of linking dual AGN to MBH mergers we do not include dual AGN at $z=0$, since by definition they cannot give rise to a MBH merger by the same redshift.} we consider first the optimistic scenario of numerical mergers, where MBHs decay rapidly from a distance of 4~kpc to coalescence, dual AGN hosted in different galaxies lead to a MBH merger by $z=0$ in 70-80 per cent of cases, with the fraction increasing for dual AGN observed in the same galaxy. The fraction reaches 90 per cent for duals in different galaxies  (a single galaxy) powered by MBHs both with mass $>10^8 \msun$ ($>10^7 \msun$): this minimizes the probability that either AGN in the dual is powered by a wandering MBH.  The fraction of successful numerical mergers also increases, for duals in both one or two galaxies, for high  ($>0.3$) total  or cold gas fractions: this is because the simulation includes gas dynamical friction.

Adding dynamical delays levels the difference between duals in two galaxies or one, and overall decreases the probability to 30-60 per cent, increasing with decreasing redshift. The delay timescales include dynamical friction and binary evolution via stellar hardening, torques in a circumbinary disc and gravitational wave emission. Dynamical friction is based on the stellar component of the galaxy that has been shown to be the dominant channel \citep{2019MNRAS.486..101P,2020ApJ...896..113L,2022MNRAS.510..531C}; see \cite{2022arXiv220111088K} for a post-processing approach that includes both stellar and gaseous dynamical friction, along with the effect of feedback. Binary evolution is generally driven by stellar hardening (see Fig. 1 in \citealt{2020MNRAS.498.2219V} and \citealt{2021ApJ...918L..15B} for a detailed analysis of the interplay between stellar and gaseous binary shrinking).  The fraction of dual AGN leading to a delayed MBH merger increases for duals hosted in massive different galaxies ($>10^{10} \msun$) and for massive dual MBHs ($>10^7 \msun$) hosted in the same galaxy. This is a simple consequence of the scalings of the implemented delays with MBH and galaxy mass (cf. Fig. 1 in Volonteri et al. 2020). Calculated delays being shorter for massive systems explains why the probability of dual AGN being precursors of MBH mergers increases at low redshift. 
Gas content enters only tangentially in the calculated delays, since it only affects the viscous timescales in circumbinary discs, which are inversely proportional to the Eddington ratios, and accretion rates on MBHs are calculated via the Bondi formalism. Since a high stellar density is the most favourable conditions for our calculated delays, dual AGN host galaxies with a low gas fraction, hence a high stellar fraction at a given mass, are more likely to give rise to a delayed MBH merger. A slight increase for the probability of successful delayed merger occurs for gas fractions less than 10 per cent, but also for gas fractions more than 30 per cent for duals in different galaxies (for which both dynamical friction in the simulation and postprocessed delays operate, since the MBHs are separated by more than $>$4~kpc, the distance below which we include the postprocessed delays). 

In analogy with the left panel, we show the effect of including/excluding dual AGN observed in the same galaxy that cannot be traced to a previous galaxy merger. Given that this population dwindles as redshift decreases, so does its effect. Furthermore, such population disappears if we consider delayed mergers since these are small MBHs that have long post-processed delays.

In the Appendix we show how several of the results presented in the paper depend on distance and luminosity cuts. We here briefly comment on the main effects and provide a more detailed discussion in the Appendix. Decreasing the distance cut to, e.g., 10~kpc increases significantly the probability that a MBH merger follows for duals hosted in two galaxies, while there is little change for duals hosted in a single galaxy. This is because most duals hosted in one galaxy are separated by less than 10~kpc, as shown in Fig.~\ref{fig:dist_same_diff_lum}. 

Increasing the dual luminosity threshold to $\log(L_{\rm bol})>44$ slightly increases the probability that a MBH merger results from the dual AGN for duals hosted in a single galaxy. This is because the most luminous duals are more centrally concentrated, and are more likely the product of a recent merger with effective dynamical friction on massive galaxies and MBHs, rather than ``wandering MBHs''. Overall, duals that give rise to merging MBHs have slightly higher Eddington ratios, but the difference is not statistically significant.

The middle panel of Fig.~\ref{fig:deltat} conveys a message similar to what has been discussed, while the bottom panel ties together galaxy and MBH mergers, especially in the case of duals in different galaxies. Although the time when a dual is observed is only a snapshot in the merger history, duals that are identified at small separations and/or have high luminosities are generally better indicators of effective mergers, which take the shortest between galaxy and MBH mergers, because of either favourable orbits or higher masses and accretion rates.

\section{Dual and multiple AGN number density}

\begin{figure}
	\includegraphics[width=\columnwidth]{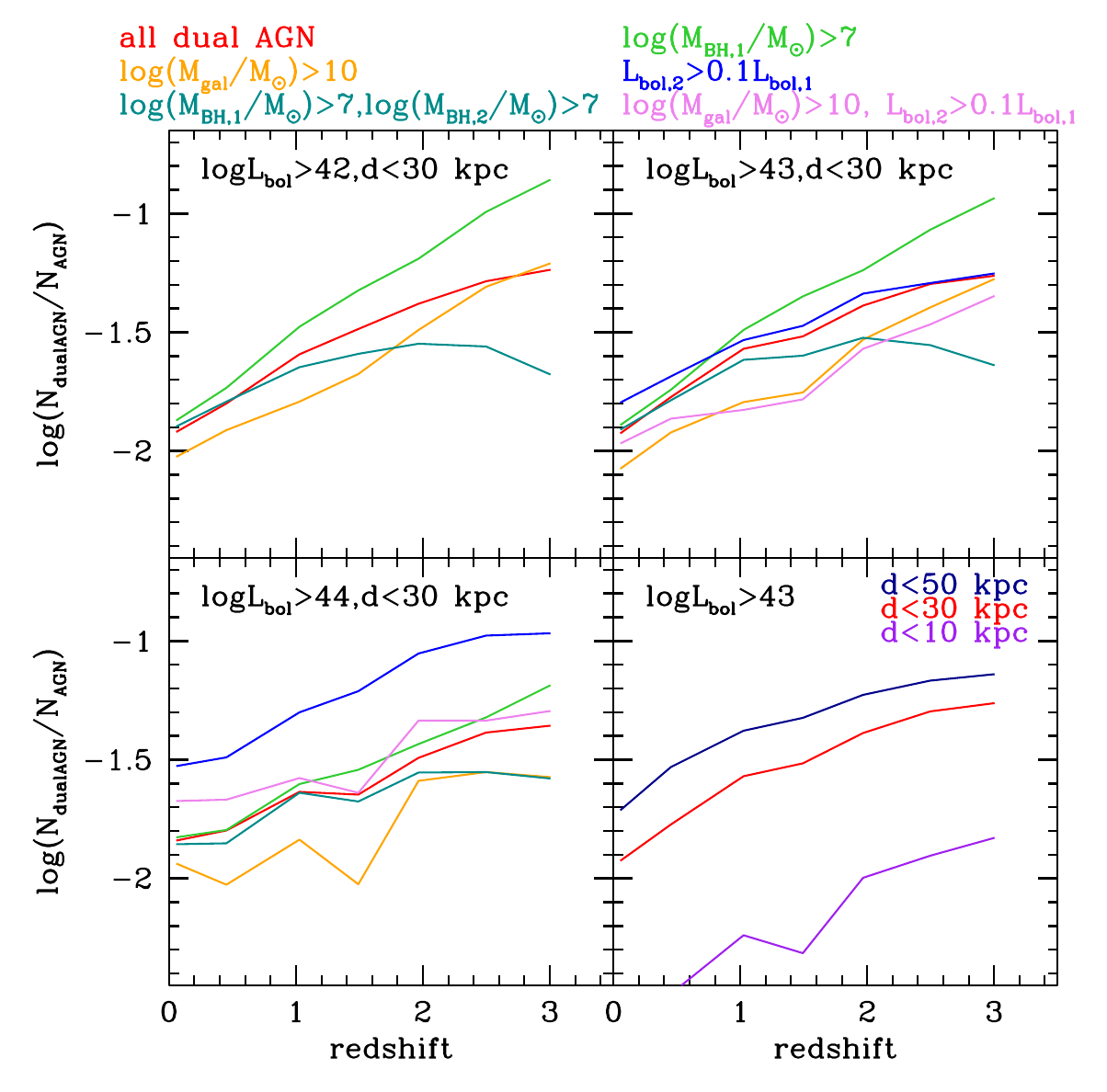}
    \caption{Fraction of dual AGN passing some threshold criteria (luminosity, BH mass, galaxy mass, distance).  Imposing mass cuts changes somewhat the overall evolution with redshift and to a higher degree the normalization. }
    \label{fig:frac_dual}
\end{figure}

\begin{figure}
	\includegraphics[width=\columnwidth]{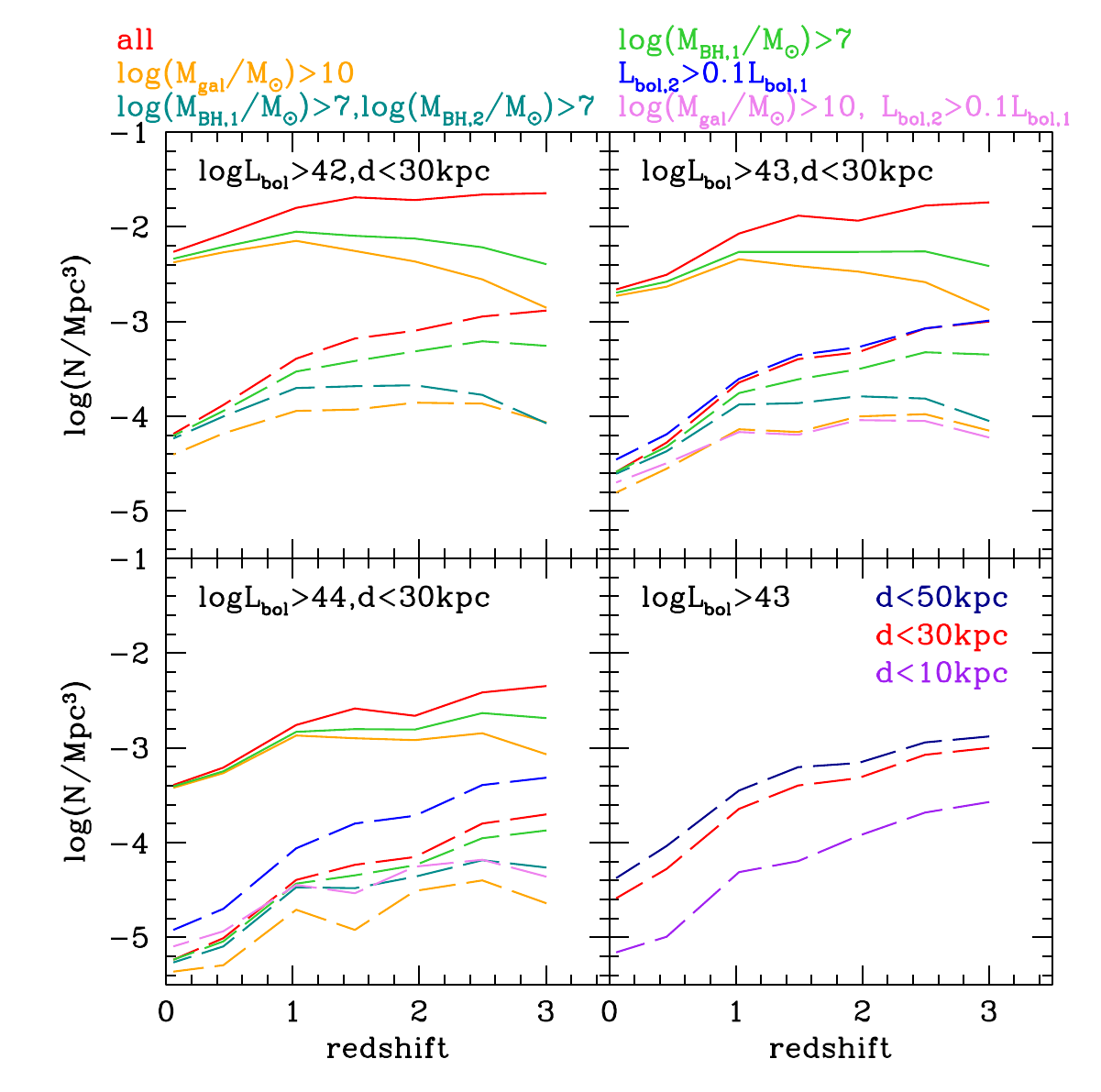}
    \caption{Number density of AGN passing some threshold criteria (luminosity, BH mass, galaxy mass; solid curves) and dual AGN passing some threshold criteria (luminosity, BH mass, galaxy mass, distance; dashed curves). As we move to higher and higher redshift, massive galaxies and MBHs dwindle, this needs to be ``convolved' with the fact that massive galaxies host duals more frequently. The dual AGN fraction can increase with redshift when imposing BH/galaxy mass cuts not because there are more duals, but because there are fewer AGN overall. This is best exemplified by the orange curves, imposing a cut in galaxy mass.}
    \label{fig:n_dualagn}
\end{figure}

\begin{figure}
	\includegraphics[width=\columnwidth]{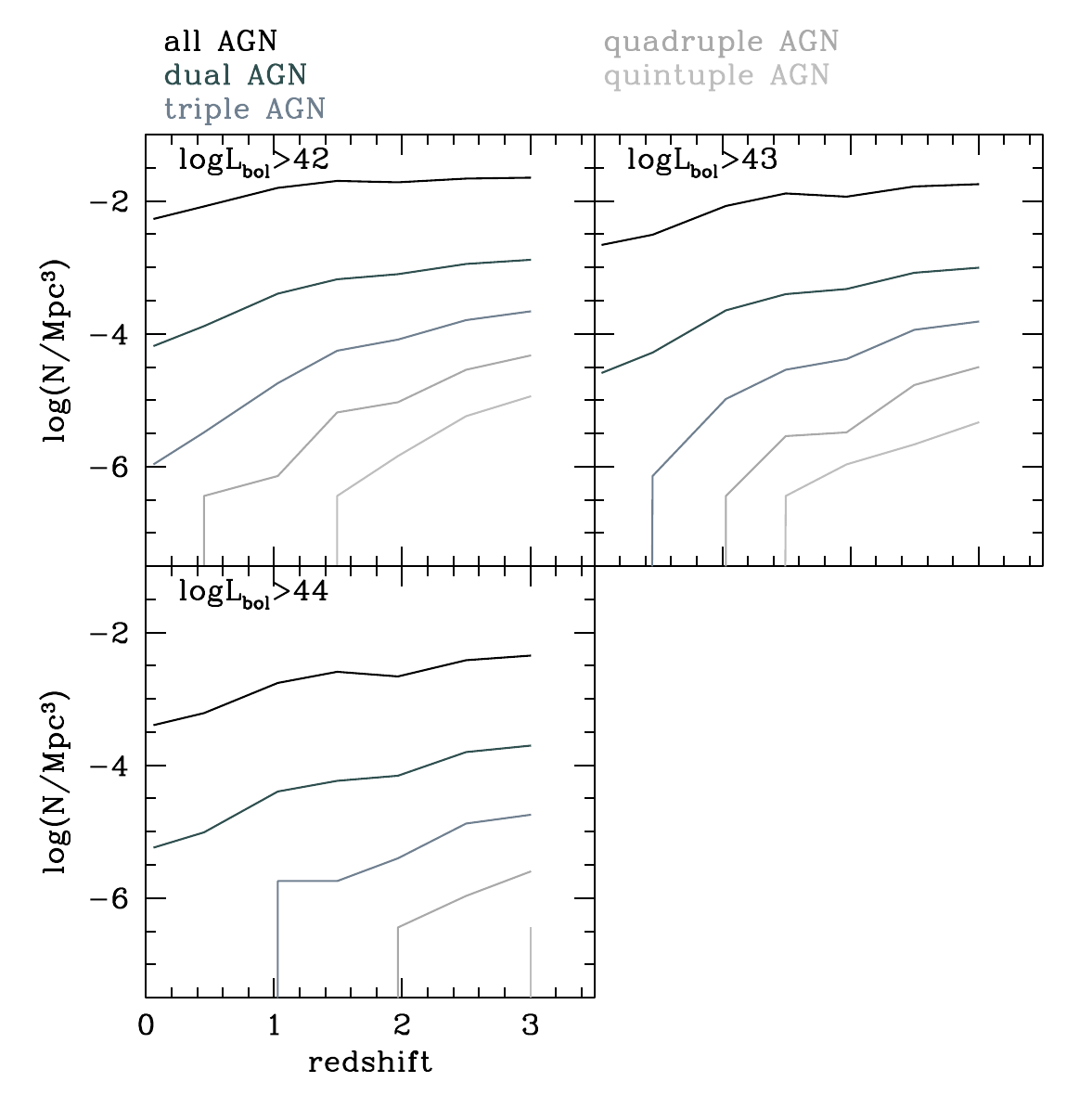}
    \caption{Redshift evolution of the number density of AGN regardless of their multiplicity (all AGN) and evolution of the number density of dual, triple, quadruple and quintuple AGN, in each case defined as being within a region of radius 30~kpc and above the bolometric luminosity quoted in each panel.Rarity increases with multiplicity.  }
    \label{fig:n_multiagn}
\end{figure}

We first focus on (pure) dual AGN, and examine their fraction over the whole population of AGN in Fig.~\ref{fig:frac_dual}. If we consider all duals where both AGN pass a simple luminosity criterion, the dual AGN fraction increases with increasing redshift, in agreement\footnote{Note that Volonteri et al. (2016), who also analyzed \hagn, considered only dual AGN hosted in the same galaxy and did not apply any distance criterion.} with previous simulations \citep[][]{2016MNRAS.460.2979V,2016MNRAS.458.1013S,2019MNRAS.483.2712R}. 

\cite{2020ApJ...899..154S}, who analyzed observations and compared to a dedicated analysis of Horizon-AGN applying their criteria, found no evolution with redshift, highlighting how subtle differences in the criteria can change the results. For instance, in the Horizon-AGN analysis performed for \cite{2020ApJ...899..154S} triples (or higher multiplets) were not removed from the dual sample, and the galaxy-MBH matching has also been improved in the present paper. As a consequence, if we apply to the pure dual AGN sample the same criteria used for \cite{2020ApJ...899..154S}: primary with $\log(L_{\rm bol})>45.3$, secondary with $\log(L_{\rm bol})>44.3$, galaxy masses $>10^{10} \msun$, MBH masses $>10^{8} \msun$, we obtain similar, but not identical results. Furthermore, the analysis in \cite{2020ApJ...899..154S} and in Horizon-AGN to mimic their sample, was limited to high luminosity AGN (primary with $\log(L_{\rm bol})>45.3$), and Horizon-AGN includes only a small number of such bright and rare AGN, given its volume, making the results dominated by small number statistics. In Fig.~\ref{fig:frac_dual} the cases closest in spirit to \cite{2020ApJ...899..154S} (lavender, orange and green-blue curves in the $\log(L_{\rm bol})>44$ panel) are those that show the least evolution with redshift, going in the direction of the results of \cite{2020ApJ...899..154S}.
 Additional examples of the evolution of the dual fraction in dependence of mass/luminosity criteria are shown in Fig.~\ref{fig:frac_dual}. For instance, at fixed luminosity, decreasing the separation from 30 to 10~kpc, decreases the fraction significantly. This is because in practice only duals hosted in one galaxy are selected. The fraction increases less when increasing the separation from 30 to 50~kpc, because at that separation many systems belong to higher multiples (see the Appendix).

The reasons for the sensitivity of the results to the criteria is both related to the sensitivity of the numerator and of the denominator in the dual AGN fraction. This is exemplified in Fig.~\ref{fig:n_dualagn}, which shows the evolution of the number density of dual AGN with redshift in comparison to all AGN. For instance, applying a MBH mass cut to both AGN in the sample makes the number density of dual AGN decrease faster with increasing redshift than the total number density of AGN powered by MBHs above the same mass threshold. This is simply a consequence of MBHs in secondary AGN at a fixed luminosity threshold being less massive than in the primaries and therefore a mass cut imposed on both AGN excludes a large number of dual AGN (see Fig.~\ref{fig:mbh_distr}). 

The number density of AGN with different multiplicity is shown in Fig.~\ref{fig:n_multiagn}. The whole population (``all AGN'') includes multiple systems, but is clearly dominated by single AGN. For fixed luminosity thresholds applied to both AGN, and no further criterion, the number density of multiple AGN increases with redshift, and the higher the multiplicity, the faster the fraction of multiple AGN with respect to all AGN increases with redshift, at least for dual-triple-quadruple AGN, where enough redshift bins are populated. For instance for $\log(L_{\rm bol})>43$ the fraction of dual AGN scales $\propto (1+z)^{0.22}$, that of triples $\propto (1+z)^{0.58}$ and that of quadruples $\propto (1+z)^{0.78}$. The redshift evolution is almost identical for $\log(L_{\rm bol})>42$ and  shallower for $\log(L_{\rm bol})>44$: $\propto (1+z)^{0.17}$, $\propto (1+z)^{0.38}$ and $\propto (1+z)^{0.51}$ for duals, triples and quadruples respectively. The dependence on the distance cut is discussed in the Appendix.

\section{Conclusions}

In this paper we have analyzed the properties of dual AGN, selected mostly via distance and luminosity criteria, in the \hagn~simulation at $z =0$, 0.5, 1, 1.5, 2., 2.5 and 3. We have generally distinguished between dual AGN that at the time of observation are in two different galaxies and those in one galaxy, since they trace two distinct phases before or after galaxy mergers. The main results are summarized in the following. 

\begin{itemize}
    \item Dual AGN represent about 4 per cent of the AGN population with the same luminosity. For separations between 4 and 30~kpc, duals hosted in a single galaxies are about 15 per cent of all duals. These one-galaxy duals typically have separation $<10$~kpc.
    \item The MBH-galaxy mass relation of dual AGN is consistent with that of the general AGN population, except for some secondary AGN in dual one-galaxy systems, which are ``undermassive''.
    \item The differences between dual AGN and the general AGN population have low statistical significance, but the trends are as follows. Primary AGN in duals are accreting at slightly higher Eddington ratios, and preferentially reside in more massive galaxies, than the general AGN population. Secondary AGN have Eddington ratios similar to, or slightly smaller than, the general AGN population; their host galaxies are compatible with those of the general AGN population, although marginally more massive. However, the AGN hosted in the smaller galaxy has generally a higher Eddington ratio than the AGN hosted in the larger galaxy. 
   \item Multiple AGN are generally associated with massive halos, with halo mass increasing with multiplicity.  The galaxy/halo mass ratios of multiple AGN present significant tails caused by mass loss of satellites in the potential of the main halo. 
    \item The vast majority of dual AGN can be associated to a galaxy merger, with a typical mass ratio of 0.2. Mass loss of the smaller galaxy and star formation in the larger galaxy during the merger decrease the mass ratio as the merger progresses.
    \item Depending on the assumptions on MBH dynamics, between 30 and 80 per cent of dual AGN with separations between 4 and 30~kpc can be associated to an ensuing MBH merger. 
    \item The dual AGN fraction increases with redshift, except for systems hosted in massive galaxies/powered by high mass MBHs. The fraction of higher multiple AGN increases with redshift at a faster rate the higher the multiple.
    \item Increasing the separation threshold for dual AGN selection does not increase the fraction proportionally, because more systems become classified as multiple AGN rather than duals.
\end{itemize}

The dual and multiple AGN catalogs generated in this study are made publicly available to ease comparison with other simulations and observations. We stress that small differences in how dual AGN are selected can lead to large differences in the results. For instance, if multiple AGN are not first removed from the dual AGN catalog, dual AGN are highly overestimated because, e.g., one single triple system could be counted as up to 3 separate dual AGN. Mass cuts also play an important role in modifying the properties of the sample. Although imposing mass cuts could hide some of the underlying population properties, applying such cuts when comparing theoretical/observational samples would be worthwhile to ensure consistency. 

\section*{Acknowledgements}
MV thanks Alessandra De Rosa for helpful feedback and suggestions. This work was granted access to the HPC resources of CINES under the allocations {2013047012, 2014047012 and 2015047012} made by GENCI. Part of the analysis of the simulation was carried out using the DiRAC facility, jointly funded by BIS and STFC. This work has made use of the Horizon and Infinity Clusters hosted by Institut d'Astrophysique de Paris, we thank  S. Rouberol for running them smoothly for us.
HP acknowledge support from the Danish National Research Foundation (DNRF132) and the Hong Kong government (GRF grant HKU27305119, HKU17304821).
GM acknowledges support from a Netherlands Research School for Astronomy (NOVA), Virtual Institute of Accretion (VIA) postdoctoral fellowship.

\section*{Data Availability}

The data underlying this article were provided by the Horizon-AGN collaboration by permission. The catalogues of dual AGN generated for this article  are available in the CDS data base, via \href{ftp://cdsarc.u-strasbg.fr/pub/cats/J/MNRAS/}{ftp://cdsarc.u-strasbg.fr/pub/cats/J/MNRAS/}  or  \href{http://cdsarc.u-strasbg.fr/viz-bin/qcat?J/MNRAS/}{http://cdsarc.u-strasbg.fr/viz-bin/qcat?J/MNRAS/}.

%%%%%%%%%%%%%%%%%%%%%%%%%%%%%%%%%%%%%%%%%%%%%%%%%%



\bibliographystyle{mnras}
\bibliography{biblio} % if your bibtex file is called example.bib

%%%%%%%%%%%%%%%%%%%%%%%%%%%%%%%%%%%%%%%%%%%%%%%%%%

%%%%%%%%%%%%%%%%% APPENDICES %%%%%%%%%%%%%%%%%%%%%

\appendix
\label{appendix}

\section{Dependence on luminosity and distance}
\label{app}

\begin{figure*}
	\includegraphics[width=\columnwidth]{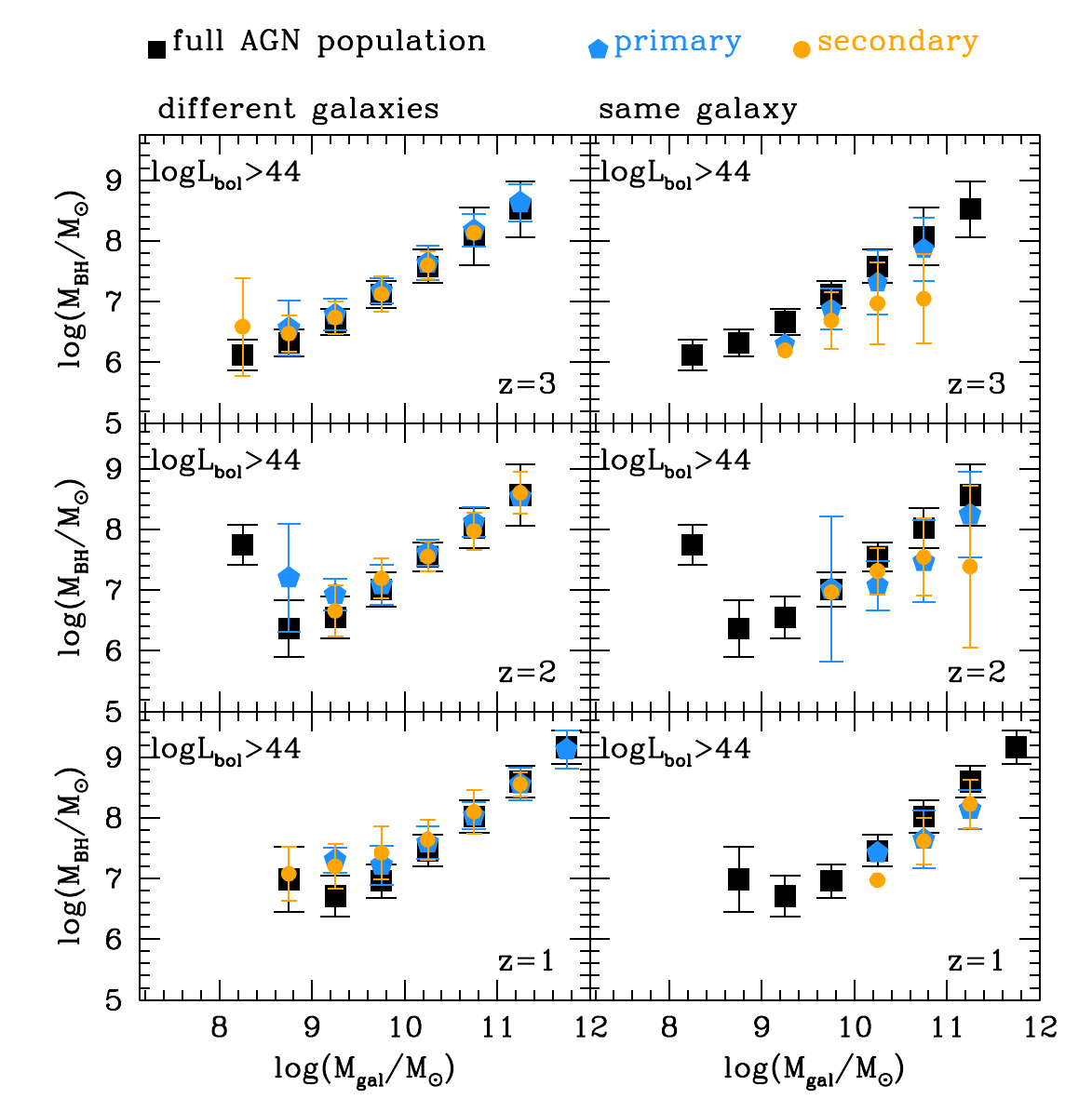}
	\includegraphics[width=\columnwidth]{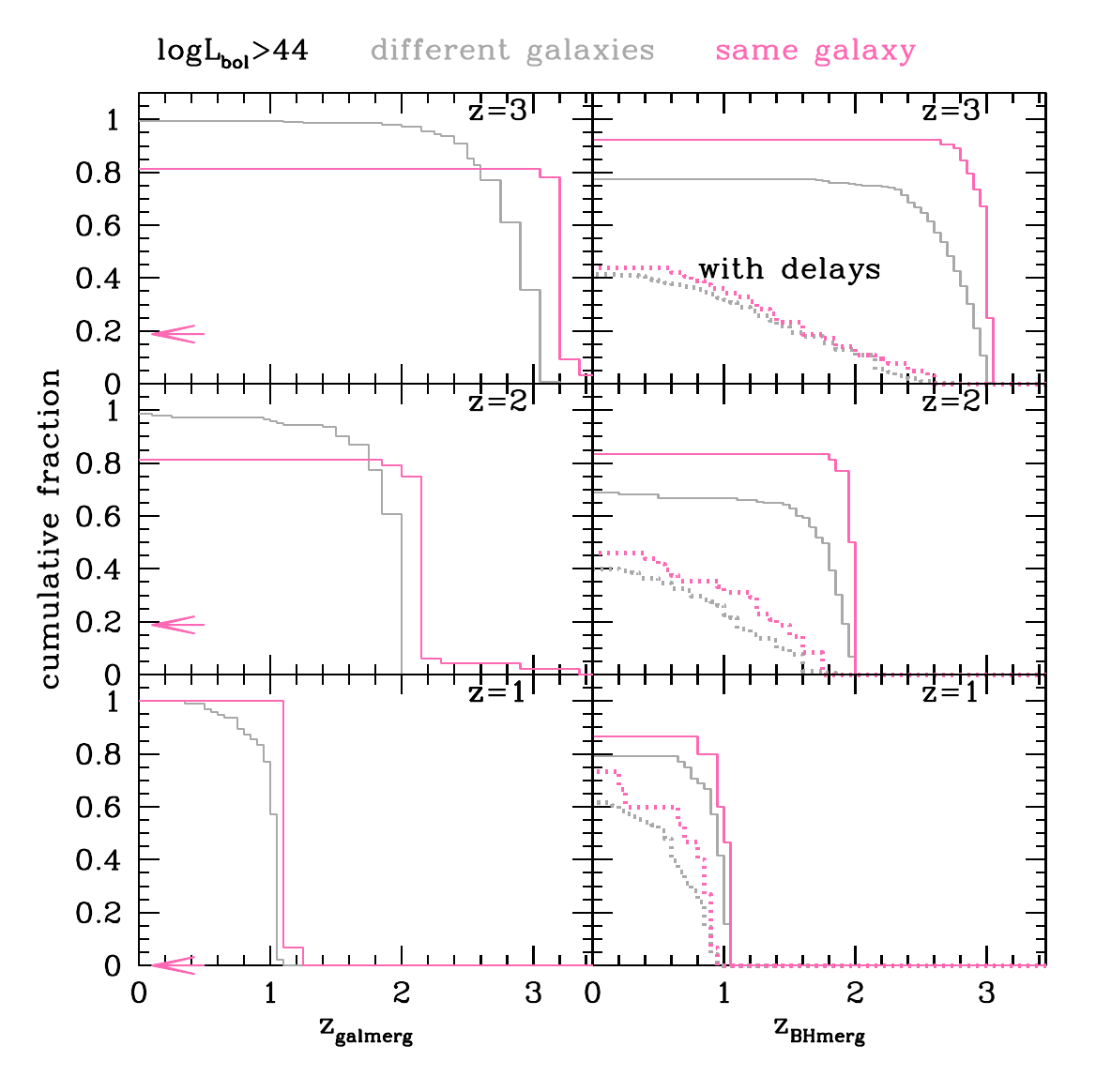}
    \caption{Analogues of Figures~\ref{fig:mbh_mgal},~\ref{fig:zdual_zBHmerge_zgalmerge_diffgal}, but for a luminosity threshold of $\log(L_{\rm bol})>44$.} 
    \label{fig_app:44}
\end{figure*}

When increasing the luminosity threshold -- from  $\log(L_{\rm bol})>43$ to $\log(L_{\rm bol})>44$ -- for dual selection the sample becomes smaller (1490 instead of 8306 objects) and therefore more prone to small number statistics. The relation between MBH and galaxy mass remains similar, with the only difference that MBHs powering secondary AGN at high redshift are closer to the general relation, i.e., less ``undermassive'', for the ``same galaxy'' case (Fig.~\ref{fig_app:44}). 

\begin{figure*}
	\includegraphics[width=\columnwidth]{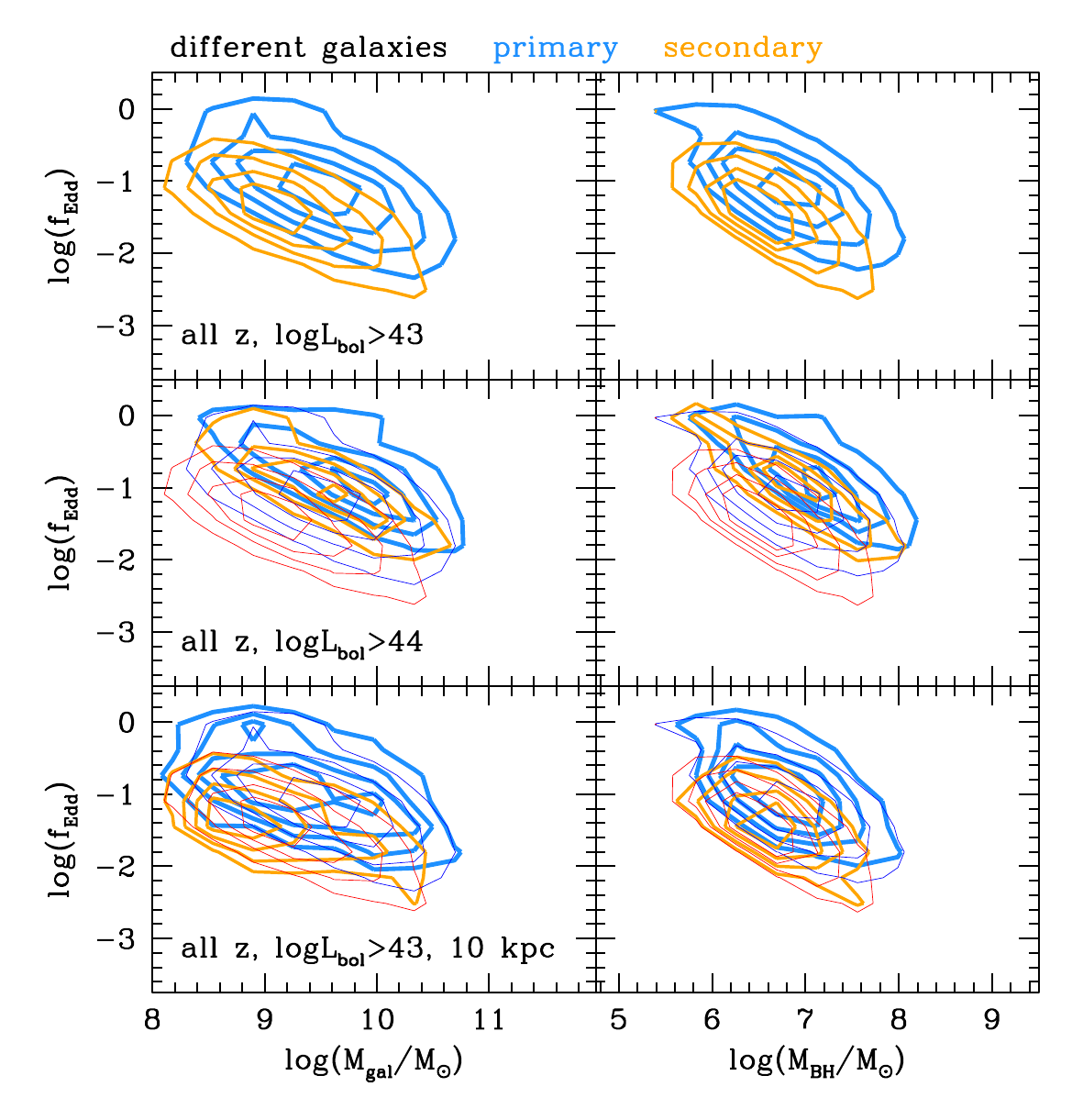}
		\includegraphics[width=\columnwidth]{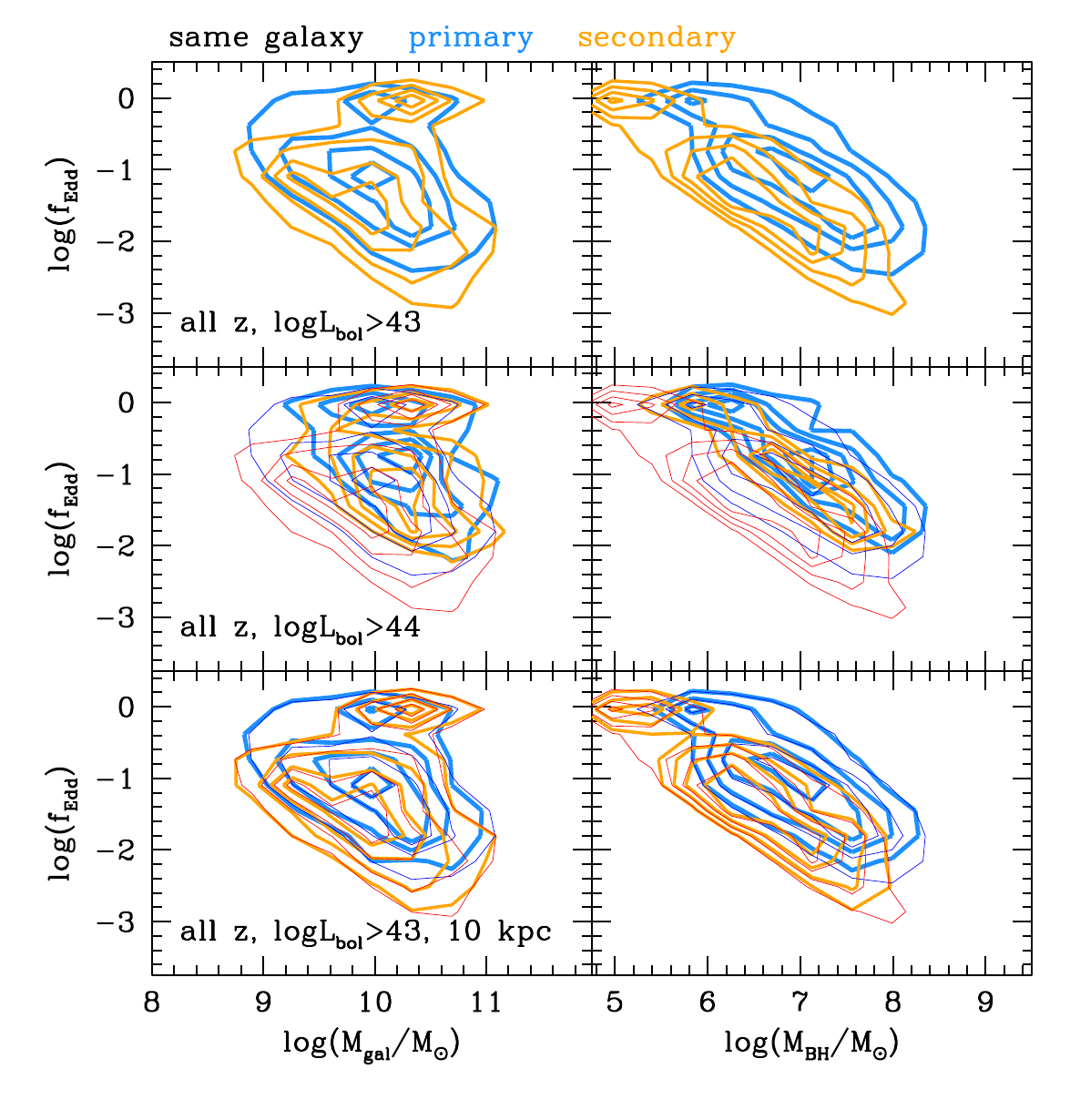}
    \caption{Similarly to Fig.~\ref{fig:fedd_mass}, the Eddington ratio as a function of galaxy and MBH mass for for primary (thick blue contours) and secondary (medium thickness orange contours) AGN, using different luminosity/distance thresholds, as marked in the panels, distinguishing dual AGN hosted in different galaxies (left) and in the same galaxy (right). The $\log(L_{\rm bol})>43$ results are reported in the middle/bottom panels as thin dark/blue (primary) and red (secondary) contours to guide the eye.}
    \label{fig_app:fedd_mass}
\end{figure*}

For duals hosted in different galaxies increasing the luminosity cut increases mostly the Eddington ratio, while for duals hosted in one galaxy an increase in the mass of the MBH powering the secondary AGN is also evident (Fig.~\ref{fig_app:fedd_mass}). High luminosity duals hosted in one galaxy have smaller separations, and a shorter delay between galaxy and MBH merger. A high luminosity threshold weeds out wandering MBHs, by selecting either more massive MBHs or MBHs in dense regions, either way dynamical friction is more efficient, so we are selecting duals on the way to merger rather than wandering MBHs. For high luminosity duals hosted in different galaxies the luminosity threshold does not make a large difference, the masses of secondary AGN are similar and \fedd~is not as close as unity as in the same galaxy case;  furthermore \fedd~decreases  as mass increases, so at high luminosity we may pick either a massive MBH in a low density environment or a light MBH accreting at high rates, in either case the orbital decay is inefficient. 

The opposite is true if the luminosity threshold is decreased to $\log(L_{\rm bol})>42$: the mass and Eddington ratio of the secondary AGN decrease, as more wandering MBHs enter the selection. A larger fraction of ``undermassive'' MBHs in secondary AGN is  for the ``same galaxy'' case is also present and the fraction of dual AGN connected to MBH mergers decreases.

\begin{figure*}
	\includegraphics[width=\columnwidth]{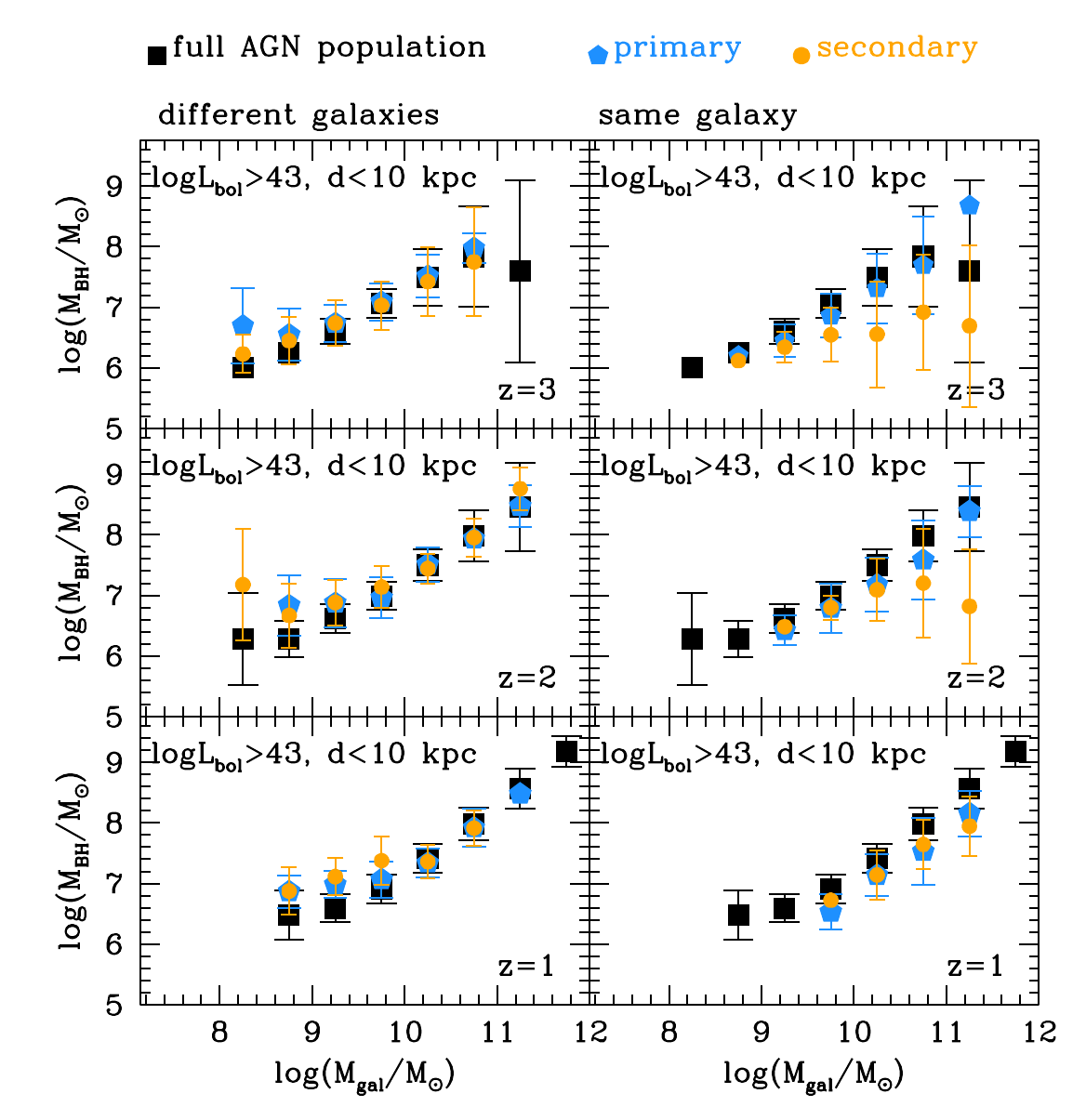}
		\includegraphics[width=\columnwidth]{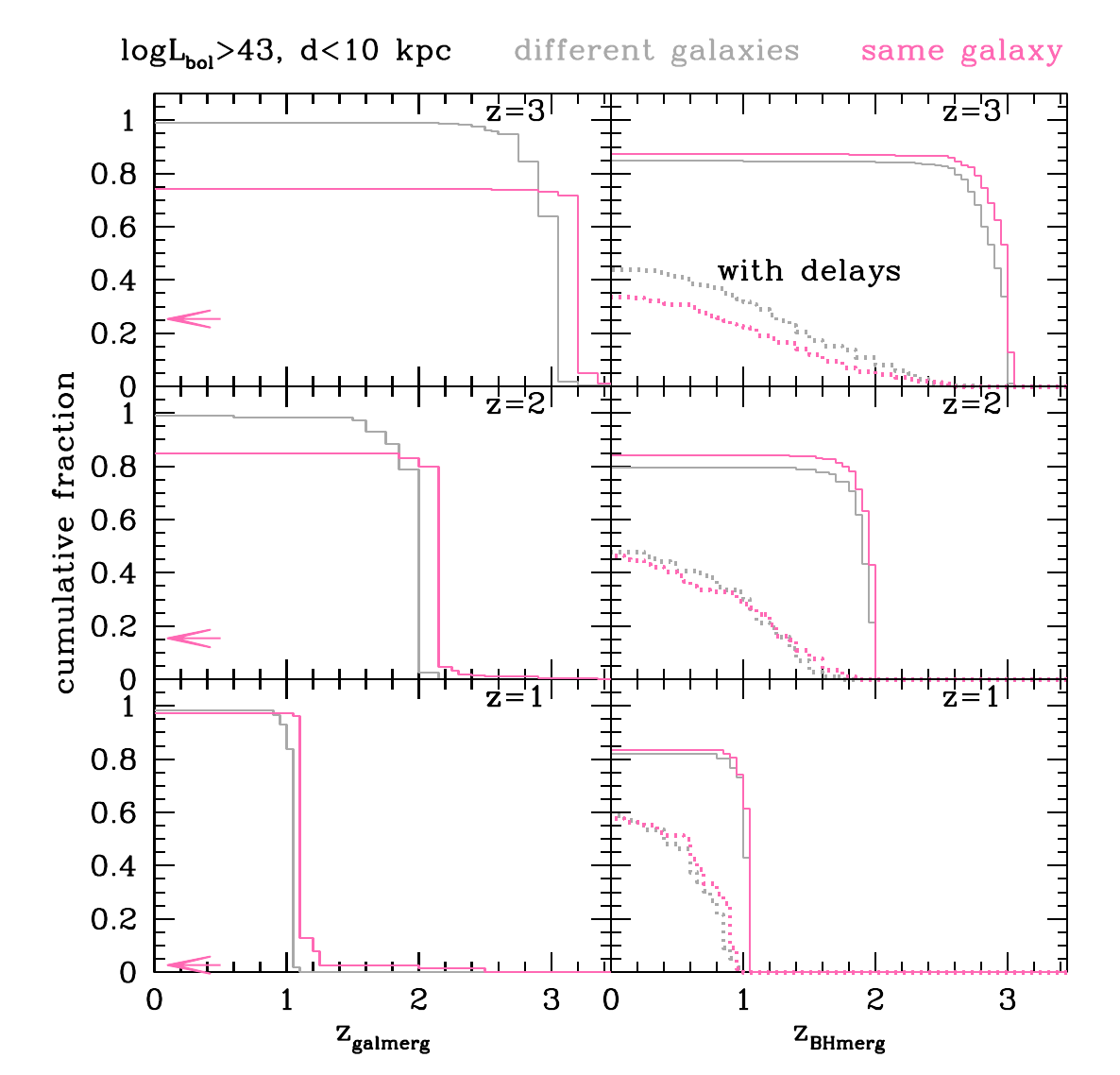}
    \caption{Analogues of Figures~\ref{fig:mbh_mgal},~\ref{fig:zdual_zBHmerge_zgalmerge_diffgal}, but for a distance threshold of $<10$~kpc.}
    \label{fig_app:dist}
\end{figure*}

When decreasing the distance threshold for dual selection -- from 30 kpc to 10 kpc --  the  sample includes 1971 objects. The relation between MBH and galaxy mass remains similar, with the only difference that MBHs for duals in different galaxies have higher mass at the low-mass end. This is likely an effect of being close to a galaxy merger: mass loss through tidal effects and difficulties in identifying all galaxy stars when two galaxies are merging both contribute to decrease the galaxy mass (Fig.~\ref{fig_app:dist}). The relationship between galaxy mass ratio and Eddington ratio does not show any statistical difference for duals hosted in either one or two galaxies (Fig.~\ref{fig_app:fedd_mass}). The increase in successful galaxy and MBH mergers is simply caused by the spatial proximity and favourable orbits.

\begin{figure*}
	\includegraphics[width=\columnwidth]{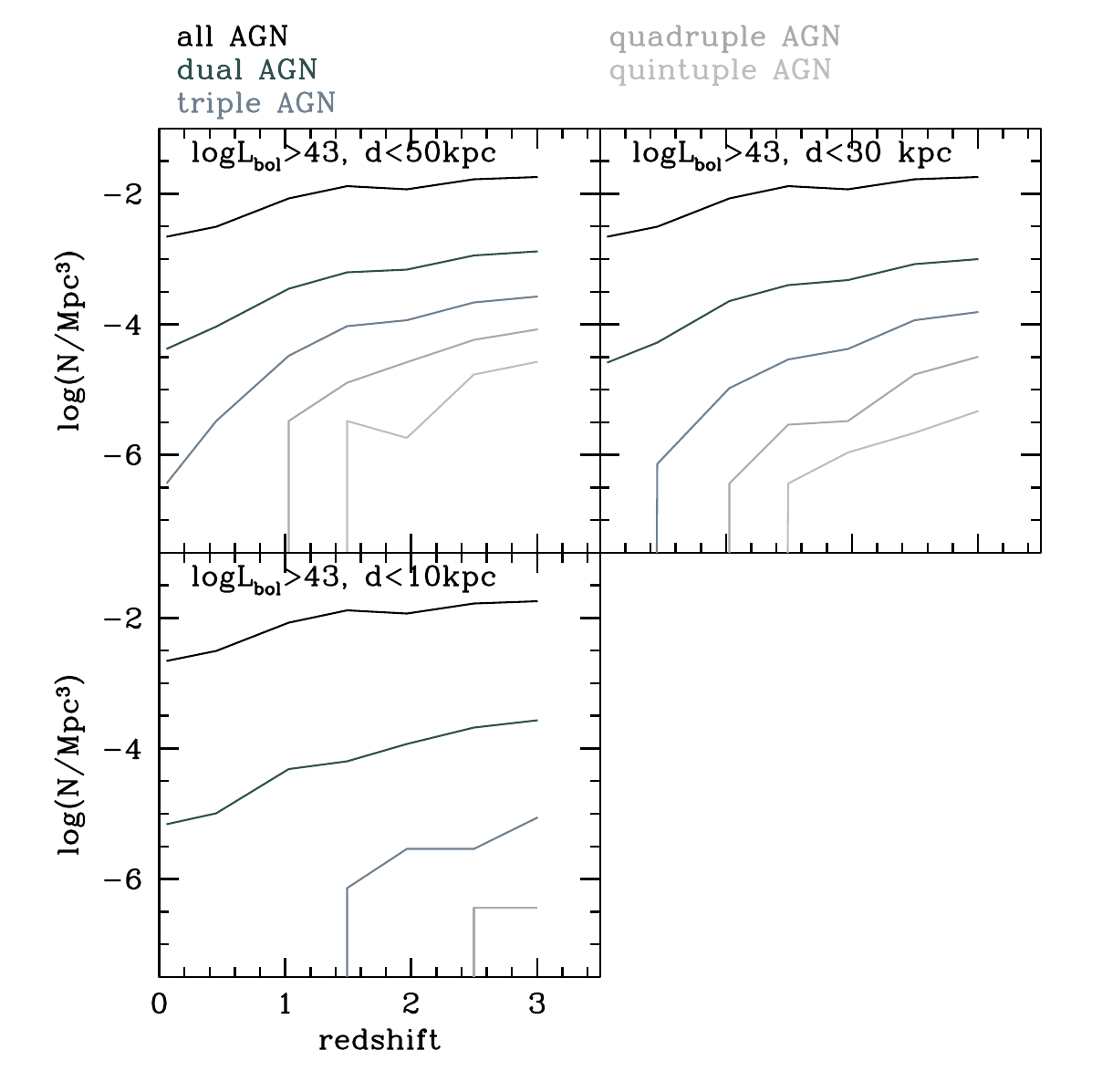}
    \caption{Analogue of Fig.~\ref{fig:n_multiagn}, but for different distance cuts.}
    \label{fig_app:n_multiagn_dist}
\end{figure*}

Finally, in Fig.~\ref{fig_app:n_multiagn_dist}, we show the redshift evolution of the number density of multiple AGN for the same luminosity threshold, but for different distance cuts out to 50~kpc. The slope of the redshift dependence does not depend much on the distance cut, the main change is in the normalization, which obviously increases as the distance cut increases.  

%%%%%%%%%%%%%%%%%%%%%%%%%%%%%%%%%%%%%%%%%%%%%%%%%%

% Don't change these lines
\bsp	% typesetting comment
\label{lastpage}
\end{document}